\documentclass[useAMS]{mn2e}

\usepackage{times}
\usepackage{amssymb}
\usepackage{rotating}
\usepackage{graphicx}
\usepackage{longtable}
\usepackage{array}
\usepackage{multirow}
\usepackage{tabularx}
\usepackage{caption}
\usepackage{epstopdf}
\usepackage{amsmath} 
\usepackage{color}
\usepackage{undertilde}
 \usepackage{siunitx}

\title[Modelling the cross-spectral variability of MAXI J1659-152]{Modelling the cross-spectral variability of the black hole binary MAXI J1659-152 with propagating accretion rate fluctuations}
\author[S. Rapisarda, A. Ingram, M. Kalamkar, M. van der Klis]{S. Rapisarda$^1$, A. Ingram$^1$, M. Kalamkar$^2$, M. van der Klis$^1$
\\
$^{1}$Anton Pannekoek Institute for Astronomy, University of Amsterdam, Science Park 904, 1098XH Amsterdam, Netherlands\\
$^{2}$INAF, Osservatorio Astronomico di Roma, Via Frascati 33, I-00078 Monteporzio Catone, Italy\\
}
\date{Accepted for publication in MNRAS}
\pagerange{\pageref{firstpage}--\pageref{lastpage}} \pubyear{2015}

\begin{document}


\pagerange{\pageref{firstpage}--\pageref{lastpage}} \pubyear{2016}

\maketitle
\topmargin = -0.5cm
\label{firstpage}

\begin{abstract}
The power spectrum of the X-ray fluctuations of accreting black holes often consists of two broad humps. We quantitatively investigate the hypothesis that the lower frequency hump originates from variability in a truncated thin accretion disc, propagating into a large scale-height inner hot flow which, in turn, itself is the origin of the higher frequency hump. We extend the propagating mass accretion rate fluctuations model \textsc{propfluc} to accommodate double hump power spectra in this way. Furthermore, we extend the model to predict the cross-spectrum between two energy bands in addition to their power spectra, allowing us to constrain the model using the observed time lags, which in the model result from both propagation of fluctuations from the disc to the hot flow, and inside the hot flow. We jointly fit soft and hard power spectrum, and the cross-spectrum between the two bands using this model for 5 \textit{Swift X-ray Telescope} observations of MAXI J1659-152. The new double hump model provides a better fit to the data than the old single hump model for most of our observations. The data show only a small phase lag associated with the low frequency hump. We demonstrate quantitatively that this is consistent with the model. We compare the truncation radius measured from our fits with that measured purely by spectral fitting and find agreement within a factor of two. This analysis encompasses the first joint fits of stellar-mass black hole cross-spectra and power spectra with a single self-consistent physical model.
\end{abstract}


\section{Introduction}
\label{sec:int}

Transient black hole X-ray binaries (BHBs) evolve in very characteristic ways during their outbursts (e.g. Belloni et al. 2005; Remillard \& McClintock 2006; Belloni 2010; Gilfanov 2010). A typical BHB outburst passes through a number of different states, each state being defined by particular spectral and timing properties of the source. At the beginning of the outburst, the source is in the low-hard state (LHS): it shows high aperiodic variability ($rms$ $\utilde{>}$ 30\%) and its energy spectrum is dominated by a hard power law component (photon index $\Gamma \approx$ 1.7). As the source luminosity increases, the source moves towards the high-soft state (HSS): the aperiodic variability drops off ($rms \approx$ 3\%), the power law softens ($\Gamma \approx$ 2.4), and the spectrum becomes dominated by a multi-colour blackbody component peaking in soft X-rays ($\approx$ 1 keV). At the end of the outburst, the source hardens again, turning back in the LHS. \\
Looking at the power spectrum of the source during the outburst, it is possible to identify several different components representing rapid variability on time scales between $\approx$ 0.01 and $\approx$ 100 s, which have different characteristics for each state. In particular, the LHS is usually characterized by the presence of a quasi periodic oscillation (QPO) superimposed on broad band continuum noise. During the evolution of the outburst, all the characteristic frequencies of the power spectral components correlate with hardness (e.g. Wijnands \& van der Klis 1998; Psaltis, Belloni \& van der Klis 1999; Homan et al. 2001). The initial transition between LHS and HSS usually takes place through intermediate states with spectral and timing properties in between those of LHS and HSS. For example, after the LHS, the source can enter the hard-intermediate state (HIMS) where its spectrum is characterized by the presence of both a disc and a power law component, the aperiodic variability decreases to $rms \approx$ 10-20\%, and the QPO superimposed on the broad band noise is still present.\\
The transition between LHS and HSS can be explained considering two different emitting regions in the accreting flow interacting with each other: an optically thick disc producing the blackbody emission (Shakura \& Sunyaev 1973), and an optically thin Comptonizing region producing the power law (Thorne \& Price 1975; Sunyaev \& Truemper 1979). The latter is often referred to as \textit{corona} (e.g. Melia \& Misra 1993; Svensson \& Zdziarski 1994; Churazov, Gilfanov \& Revnivtsev 2001) or \textit{flow} depending on whether the region is vertically or radially separated from the disc respectively. In particular, the \textit{truncated disc model} (e.g. Esin, McClintock \& Narayan 1997; Done, Gierli\'nski \& Kubota 2007) considers an optically thick geometrically thin accretion disc truncated at a certain radius $r_o$ and an optically thin geometrically thick hot flow extending from $r_o$ down to a radius equal or larger than the innermost stable circular orbit (ISCO). At the beginning of the outburst, the truncation radius is still relatively far from the black hole (BH) and the energy spectrum is dominated by the power law component. When the mass accretion rate increases, the truncation radius approaches the BH and the energy spectrum becomes dominated by the blackbody emission. Disc photons up-scatter in the hot flow cooling it down and, as a consequence, the power law softens. \\
Although the spectral properties of BHBs can be explained considering this two-regime accreting configuration (even though the precise way in which the disc and hot flow interact with each other is not clear), the origin of the fast variability is not fully understood, and a single model explaining both spectral and timing properties is a still matter of debate. The recently proposed model \textsc{propfluc} (Ingram \& Done 2011, 2012, hereafter ID11, ID12; Ingram \& van der Klis 2013, hereafter IK13) is based on the truncated disc model described above. Additionally, \textsc{propfluc} contains the ingredients of mass accretion rate fluctuations propagating through the hot flow, and precession of the entire hot flow caused by frame dragging close to the BH. Mass accretion rate fluctuations are generated at every radius of the hot flow and propagate towards the BH giving rise to a broad band noise component in the power spectrum (single hump power spectrum). The characteristic time scale of the noise is set by the viscous time scale in the hot flow (e.g. Lyubarskii 1997; Churazov, Gilfanov \& Revnivtsev 2001; Arevalo \& Uttley 2006). As a consequence of the propagation of the fluctuations, the time variability of the emission from every ring of the flow is correlated (with a time delay). Because the mass accretion rate fluctuations at larger radii, after propagating inward, modulate the amplitude of the fluctuations at smaller radii by multiplication, the process gives rise to the linear rms-flux relation observed in BHBs (Uttley \& McHardy 2001; Uttley, McHardy \& Vaughan 2005). Meanwhile, the Lense-Thirring (LT) precession of the entire hot flow (Stella \& Vietri 1998; Fragile et al. 2007; ID11) produces the QPO at a frequency depending on the mass distribution in the hot flow and on its radial dimension. \\
Rapisarda et al. (2014) (hereafter RIK14) presented the first application of \textsc{propfluc} to study the BH candidate MAXI J1543-564. They fitted selected power spectra of the rising phase of the 2011 outburst of the source with the single hump power spectrum calculated by \textsc{propfluc} and traced the evolution of the physical parameters in these observations.\\
The \textsc{propfluc} version used in RIK14 produces a single hump power spectrum originating from mass accretion rate fluctuations arising only in the hot flow. However, timing analysis of BHBs shows that their power spectrum in the LHS/HIMS is often characterized by a more complex structure than a single hump (e.g Belloni et al. 1997; Homan et al. 2001; Kalamkar et al. 2015a), requiring two or three broad Lorentzians to be fitted (low, mid, and high frequency Lorentzian). Additionally, BHBs often show time lags between different energy bands associated with this broad band variability (e.g., Miyamoto et al. 1988; Nowak, Wilms \& Dove 1999a). The delay between emission in different energy bands depends on the geometry of the accreting system and can be used to constrain different accretion models (e.g. Miyamoto \& Kitamoto 1989; B$\rm \ddot{o}$ttcher \& Liang 1999; Misra 2000; Nowak et al. 1999b; Kotov et al. 2001; Arevalo \& Uttley 2006).\\
Combining spectral and timing analysis it is possible to obtain clues about the origin of the different power spectral components. In particular, Wilkinson \& Uttley (2009), on the basis of measurements of variability amplitudes of X-ray spectral components, suggested that low frequency noise is the result of intrinsic variability generated in the disc and propagating through the flow. By its very nature, propagation also predicts time lags between soft and hard energy  bands, but up to now these two aspects of the propagation hypothesis have never been jointly considered in a quantitative analysis. \\
As pointed out in IK13, with the model \textsc{propfluc} we can simultaneously predict these time lags, the variability amplitudes, and the coherence between energy bands by calculating power spectra at different energies, and cross-spectra between those energies. These predictions can then jointly be fitted to observed power and cross-spectra. The model can also be adapted to simulate extra disc variability and produce a two-hump power spectrum by considering mass accretion rate fluctuations generated both in the disc and in the flow, all propagating towards the BH. Fits to cross- and power spectra of BHBs in the LHS/HIMS characterized by a two-hump profile can then be attempted using observations spanning the low energy range where the disc emission is concentrated. \\
In this paper, we analyze data from MAXI J1659-152, a BH discovered in 2010 (Mangano et al. 2010; Negoro et al. 2010). During its 2010 outburst, MAXI J1659-152 followed the usual behavior observed in BHB outbursts (Mu{\~n}oz-Darias et al. 2011). Previous timing analysis of the source using the \textit{Rossi X-ray Timing Explorer} (RXTE; Jahoda et al. 1996) and \textit{Swift} (Gehrels et al. 2004) observations (Kalamkar et al. 2011; Kalamkar et al. 2015a), showed that its power spectra in the HIMS are characterized by several broad band components with characteristic frequencies between $\approx 0.001$ and $\approx 5$ Hz. We explore the hypothesis put forward by Kalamkar et al. (2015a) that some of this enhanced low frequency variability originates in the disc by performing joint fits of the power and cross-spectra of MAXI J1659-152 in the HIMS using \textit{Swift} XRT data in two different energy bands (0.5 - 2.0 keV and 2.0 - 10.0 keV). The \textit{Swift} XRT data allow us to study the source from the beginning of the outburst (RXTE started observing the source 3 days later) in an energy range where the disc emission is significant.\\
Sec. \ref{sec:mod} is dedicated to the description of the new two-hump version of \textsc{propfluc}, Sec. \ref{sec:obs} briefly describes how we reduced and analyzed the data, and in Sec. \ref{sec:res} and Sec. 5 we present and discuss the results of our fits, respectively. By a strictly quantitative analysis, we find, perhaps counterintuitively, that the small lag observed in the broadband noise between these two energy bands is entirely consistent with mass accretion rate fluctuations in the disc propagating to the hot flow.

\section{The new \textsc{propfluc} model}
\label{sec:mod}

\textsc{propfluc} (ID11, ID12, IK13) is a model assuming a truncated disc/hot flow geometry, with mass accretion rate fluctuations propagating through a precessing hot flow. Here, we introduce two extra features to the model: 1) we consider that variability can also be generated in the disc, which then propagates to the hot flow on a viscous infall time; 2) we improve the model so that it is now possible to simultaneously fit power spectra in two different energy bands, and the complex cross-spectrum between these two bands. The cross-spectrum contains information on the power in each of these bands, the phase lags between the bands, and also the coherence between the bands. In order to fit the cross-spectrum, we introduce a formalism to include arbitrary QPO phase lags, which allows us to concentrate on physical modeling of the broad band noise (see Appendix B).

\subsection{Propagating fluctuations and disc variability}
\label{sec:disc}

\begin{figure} 
\center
\includegraphics[scale=0.4,angle=270]{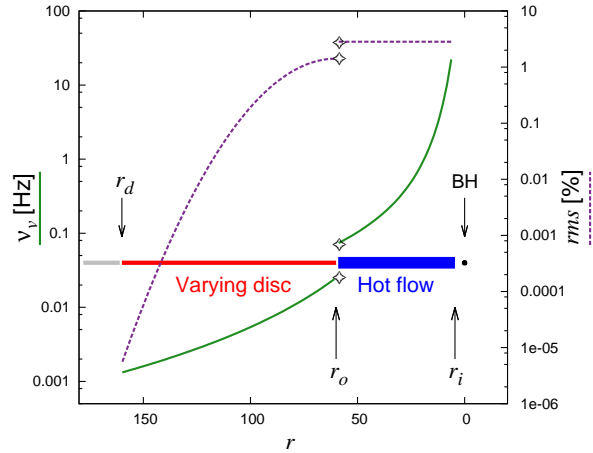}
\caption{Variability produced (purple dashed line) and viscous frequency (green solid line) versus radial coordinate $r$ for the entire mass accretion rate fluctuations propagating region (disc + hot flow). Blue and red horizontal lines indicate the hot flow and the disc respectively, the grey horizontal line ($r>160$) corresponds to the disc region that does not contribute to the variability. Both viscous frequency and amount of variability produced by every ring, are discontinuous at the truncation radius $r_o$ (star symbols).}
\label{fig:nurms}
\end{figure}

\begin{figure} 
\center
\includegraphics[scale=0.45,angle=270]{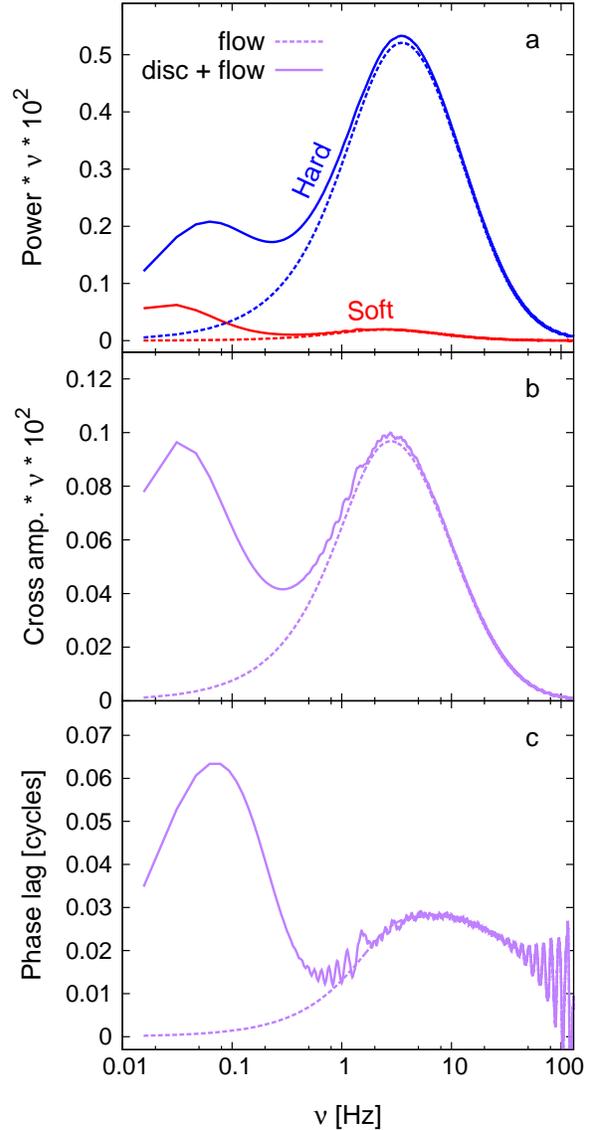}
\caption{Soft (red line) and Hard (blue line) power spectrum (a), cross spectrum (b), and phase lag (c) computed considering mass accretion rate fluctuations propagating only in the hot flow (dashed line) and in the hot flow + disc (solid line). In the second case $\textsc{propfluc}$ produces a two-hump power spectrum with an evident hard lag associated with the low frequency hump.}
\label{fig:twop}
\end{figure}

The accreting region producing the variability extends from an inner radius $r_i$ equal or larger than the innermost stable circular orbit ($r_{ISCO}$) up to a radius $r_d$ much smaller than the outer edge of the disc (in this paper we use the convention that lowercase $r$ corresponds to radial coordinate scaled by gravitational radius: $r = R/{R_{g}}$, where $R_g = GM/c^2$). The truncation radius $r_o$ is in between $r_i$ and $r_d$, dividing the region into hot flow ($r_i < r < r_o$) and varying disc ($r_o < r < r_d$, hereafter just ``disc''). The geometry of the accreting region ($r_d > r_o > r_i$) is sketched in Fig. \ref{fig:nurms}: red and blue horizontal thick lines indicate disc and hot flow respectively. From the computational point of view, the hot flow is split into rings that are equally logarithmically spaced, so that $dr/r = constant$ for each ring, where $r$ and $dr$ are radial coordinate and thickness of the ring respectively. The power spectrum of mass accretion rate fluctuations generated in each ring of the hot flow is a zero-centered Lorentzian with the width set by the local viscous frequency $\nu_{v,flow} (r)$ and having amplitude $\sigma_{0}^2 = ( F_{var} / \sqrt{N_{dec}} )^2$, where the model parameters $F_{var}$ and $N_{dec}$ are the fractional variability produced per radial decade and the number of rings per radial decade, respectively. The right part of Fig. \ref{fig:nurms} ($r < r_o = 60$) shows the dependence of viscous frequency (green solid line) and $\sigma_0$ (purple dashed line) on radius in the hot flow. Whereas the variability amplitude is assumed to be constant within the hot flow, the viscous frequency in there is described by a smoothly broken power-law (see Eq. 2 and 3 in ID12). This derives from our assumption that the surface density in the hot flow is characterized by a smoothly broken power-law, since, from mass conservation, the viscous frequency in every ring is inversely proportional to the average surface density (Frank, King, \& Raine 2002). The power spectrum of the emission produced by mass accretion rate fluctuations generated in each ring and propagating through the hot flow towards the BH, is a broad band component (single ``hump'' power spectrum) with approximately constant $rms$ (i.e. power $P(\nu) \propto 1/{\nu}$) between low and high frequency breaks, where the low frequency break is the viscous frequency at the outer edge of the hot flow $\nu_{v,flow} (r_o)$ and the high frequency break depends on the highest frequency produced in the hot flow, but it is also influenced by coherent addition of the variability emitted from different regions of the hot flow.  \\
The disc is also split into rings that are equally logarithmically spaced, and the number of rings per radial decade, i.e. the model radial resolution, is the same as in the hot flow ($N_{dec,flow} = N_{dec,disc} = N_{dec}$). The power spectrum of mass accretion rate fluctuations generated in each ring within the disc region is also characterized by a zero-centered Lorentzian with width set by the local viscous frequency. The left part of the plot in Fig. \ref{fig:nurms} ($r_o = 60 < r < r_d = 160$) shows the dependence of viscous frequency (green solid line) and variability amplitude (purple dashed line) on radius for the disc. The variability amplitude is assumed to peak at the inner edge of the disc, $r_o$, and drop off outside of this with a Gaussian dependence on radius. The peak and the width of this Gaussian, $\sigma_0 N_{var} $ and $\Delta d$, respectively, are both model parameters. We set $r_d = r_o + 5 \Delta d$, since to a very good approximation there is no variability outside this radius. For the disc we assume the viscous frequency profile of a Shakura-Sunyaev disc with constant viscosity parameter and scale-height (Shakura \& Sunyaev 1973):
\begin{equation}
\nu_{v, disc} (r)= \nu_{d, max} ( r / r_o )^{-3/2}
\label{eq:nudisc}
\end{equation}
where the viscous frequency at the inner edge of the disc, $\nu_{d, max} \equiv \nu_{v,disc}(r_o)$, is a model parameter. Mass accretion rate fluctuations generated in every ring, from $r_i$ to $r_d$, propagate towards the BH on a viscous infall time. We see that the model allows for a discontinuity in the viscous frequency at the truncation radius as the accretion flow transitions from disc to hot flow (see Fig. \ref{fig:nurms}, star symbols). It is this jump in frequency that results in two humps in the predicted power spectrum, with the lower frequency hump contributed by the disc (since the viscous frequency is lower here) and the higher frequency hump contributed by the hot flow.\\
As with previous versions of the model, we assume that the count rate observed in a given energy band can be represented as a linear combination of the mass accretion rate in each ring. The count rate in a ``hard band", $f_h(t)$, is given by:
\begin{equation}
f_h(t) = \sum_{j=1}^N h(r_j) \dot{m}(r_j,t)
\end{equation}
where $N$ is the \textit{total} number of rings between $r_i$ and $r_d$ (i.e. $N$ is the number of rings in the hot flow plus the number of rings in the disc), the emissivity function $h(r_j)$ is the mean count rate observed from the $j^{th}$ ring in this energy band, and $\dot{m}(r_j,t)$ represents the varying mass accretion rate in the $j^{th}$ ring. If we know (or, rather, make an assumption for) the mean spectrum emitted from each ring and the detector response, we can directly calculate $h(r_j)$ from the counts spectrum of the $j^{th}$ ring. For the flow, in the absence of a standard model for the spectrum as a function of radius, we simply parameterize $h(r_j)$ as a power-law function of $r$ with an inner boundary condition given by the surface density profile (see Appendix A and ID12). \\
For the disc, in contrast, we do have a standard model: a blackbody with temperature $\propto r^{-3/4}$ and luminosity $\propto r^{-3}$ (Shakura \& Sunyaev 1973). To calculate each $h(r_j)$ for the disc, we start with the blackbody spectrum from radius $r_j$ (see Fig. \ref{fig:bb}), convolve it with the telescope response, and integrate over the energy range of interest (see Appendix A for details). \\
For both the disc and hot flow, we expect the spectrum to be harder for smaller $r$, which translates to a hard band emissivity function $h(r_j)$ being a steeper function of $r$ (i.e. more centrally peaked) than a soft band emissivity function $s(r_j)$. We note that our assumption of linearity relies on variability in disc temperature, $T(r,t)$, being much smaller than the variability in $\dot{m}(r,t)$. This is a good assumption, since $T \propto \dot{m}^{1/4}$. The maximum temperature reached by the disc (at $r_o$ since we do not employ the zero-torque boundary condition - see Appendix A), $T_{d,max}$, is a model parameter. \\
We also need to parameterize the normalization of the disc spectrum. The absolute normalization is not of interest to us, but the \textit{fraction} of the total photons observed in a given band that are contributed by the disc \textit{is} of interest. In this paper, we consider two energy bands, a soft band $s$ and a hard band $h$. The fraction of observed disc photons in the soft band, $x_s$, is a model parameter which, along with $T_{d,max}$, can be measured from a spectral fit. The disc fraction in the hard band, $x_h$, can be calculated from $x_s$ and the hardness ratio $HR$ (the ratio between counts in the hard and soft band), which can be measured directly from the soft and hard light curves (see Appendix A). Tab. \ref{tab:pars} lists all the new model parameters with a short description. \\
We compute the power spectrum of $f_s(t)$ and $f_h(t)$ (0.5-2.0 keV and 2.0-10 keV, respectively), and the cross-spectrum between $f_s(t)$ and $f_h(t)$ using the formulae from IK13. Fig. \ref{fig:twop}$a$ shows soft and hard power spectra produced considering variability generated in both the disc and the hot flow (solid line) and the only hot flow (dashed line). Introducing a propagating region in the disc has two evident effects: 1) the power spectrum consists of two broad band components (double hump power spectrum), a low frequency one generated in the disc and a high frequency one generated in the hot flow; 2) the total power in the disc + hot flow case is higher than the only hot flow case because of extra variability coming from the disc.\\
Due to propagation, the hot flow emission lags the disc emission. Since the disc emits a softer spectrum than the hot flow, we expect disc variability to contribute a hard phase lag (i.e. hard photons lagging soft photons). The amplitude of this phase lag depends on the viscous infall time and on how the disc and hot flow emission are distributed in the soft and hard band. If $x_s=1$ and $x_h=0$, the soft band exclusively contains disc emission and the hard band exclusively contains hot flow emission. Therefore the lag between these energy bands is equal to the lag between the two physical components. For $x_s<1$ and $x_h>0$, the lag between energy bands is diluted by a contribution by each physical component to both bands. Panels $b$ and $c$ of Fig. \ref{fig:twop} show cross-amplitude and phase lag between soft and hard band respectively; for making the plots we set $x_s = 0.8$, $x_h = 0.5$, and we use the response matrix of \textit{Swift}. In this paper, we adopt the usual convention that a positive phase lag corresponds to the hard band lagging soft. With these assumptions, mass accretion rate fluctuations propagating from the disc produce a clear positive phase lag of $\approx 0.065$ cycles in the low frequency hump (see Fig. \ref{fig:twop}$c$ between $\approx 0.01$ and $0.1$ Hz).

\subsection{\textsc{propfluc} outputs}
\label{sec:exp}

\begin{figure} 
\center
\includegraphics[scale=0.5,angle=270,trim=0cm 0 0 0,clip]{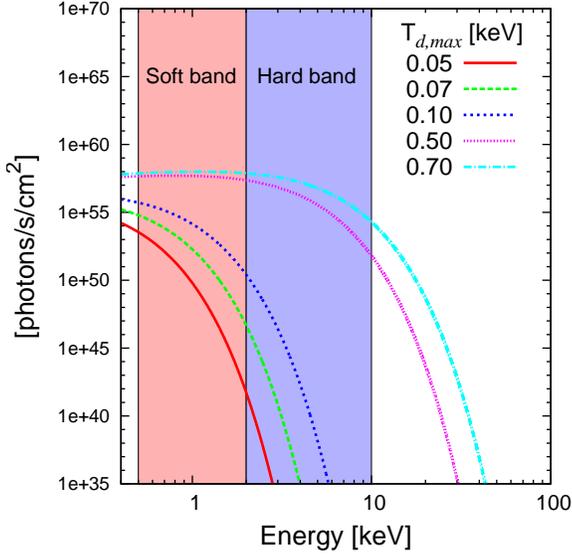}
\caption{Photon flux emitted by a single ring $r$ in the disc characterized by viscous frequency $\nu_{v, disc}(r)$ = 0.1. The red and blue regions represent soft and hard band respectively.}
\label{fig:bb}
\end{figure}

\begin{figure} 
\center
\includegraphics[scale=0.4,angle=270,trim=1cm 0 0 0,clip]{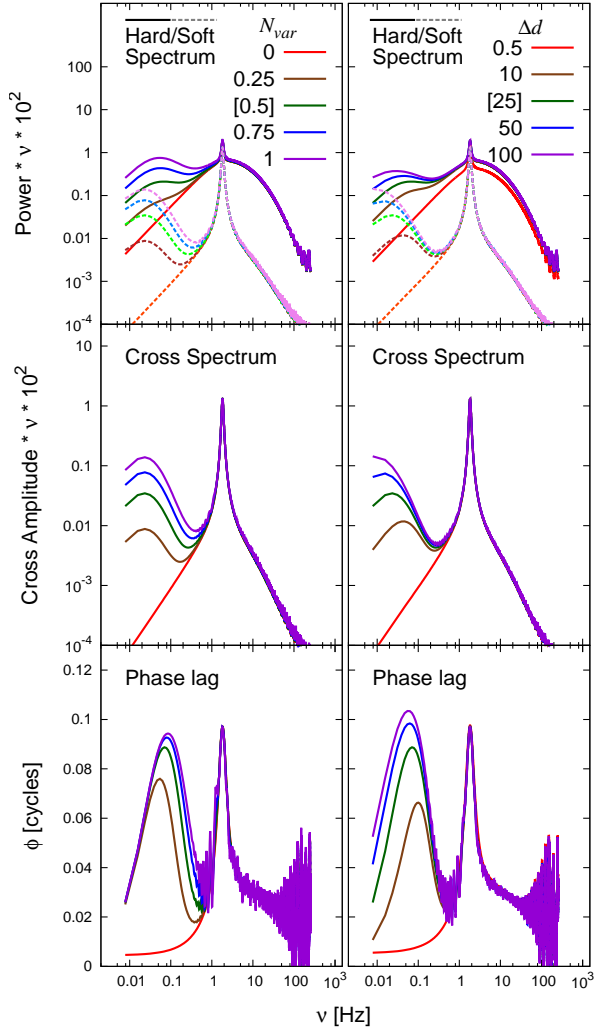}
\caption{Soft (dashed line) and hard (solide line) power spectra, cross spectra, and phase lags computed varying model parameters $N_{var}$ (left column) and $\Delta d$ (right column) as indicated. Numbers in square brackets indicate the parameter value for all the other computations.}
\label{fig:par1}
\end{figure}

\begin{figure} 
\center
\includegraphics[scale=0.4,angle=270,trim=1cm 0 0 0,clip]{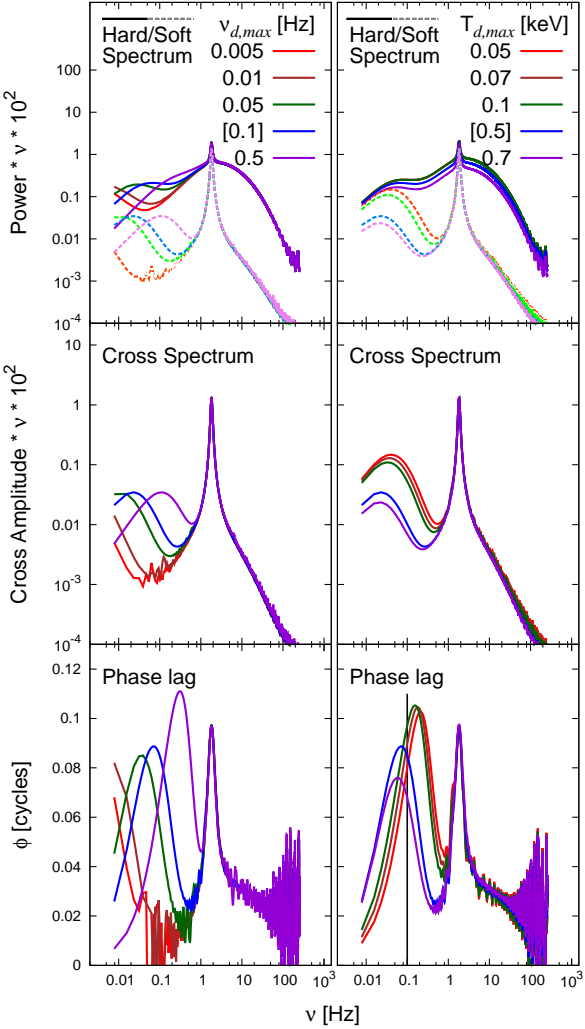}
\caption{Soft (dashed line) and hard (solide line) power spectra, cross spectra, and phase lags computed varying model parameters $\nu_{d,max}$ (left column) and $T_{d,max}$ (right column) as indicated. Numbers in square brackets indicate the parameter value for all the other computations.}
\label{fig:par2}
\end{figure}

In RIK14 we computed power spectra in a single energy band varying the model parameters related to the hot flow in order to show the relation between canonical multi-Lorentzian fitting parameters and \textsc{propfluc} parameters. Here, we compute soft and hard power spectra, and cross-spectra (implying the phase lags and coherence) between the bands changing the model parameters related to the disc (Tab. \ref{tab:pars}): the maximum viscous frequency in the disc $\nu_{d,max}$ (see Eq. \ref{eq:nudisc}), the radial extension of the disc $\Delta d$, the variability produced in every ring of the disc parameterized as fraction of the hot flow variability $N_{var}$, and the maximum temperature in the disc $T_{d,max}$. We fix all the parameters related to the hot flow, which are already discussed in RIK14: the surface density constant ($\Sigma_0 = 6$), the smoothly broken power law describing the surface density profile ($\kappa = 3.0$, $\lambda = 0.9$, $\zeta = 0$), the inner radius ($r_i = 4.5$), the transition radius of the smoothly broken power law ($r_{bw} = 7.0$), the truncation radius ($r_o = 20$), the fractional variability ($F_{var} = 0.3$), the soft and hard band emissivity indices ($\gamma_s = 3.0$, $\gamma_h = 4.5$), the BH mass ($M = 10M_{\odot}$), and the dimensionless spin parameter ($a_{*}=0.5$). We compute soft (0.5-2.0 keV) and hard (2.0-10 keV) power spectra, and cross-spectra between soft and hard band, with a Nyquist frequency of 128 Hz, using a model resolution of $N_{dec} = 35$, fixing the disc fraction in the soft band ($x_s = 0.9$), the hardness ratio ($HR = 1.0$), and including a main QPO with fixed width, $rms$, and phase lag ($Q=8$, $\sigma_{qpo} = 5\%$, $\phi_{QPO}$ = 0.1 cycles). We did not include any other QPO harmonic component for simplicity (in Appendix \ref{qpo} we describe the details of including the QPO in the new \textsc{propfluc} model). We computed all the timing products taking into account the \textit{Swift} response matrix and we considered interstellar absorption with a column density of $n_H = 1.7 \times 10^{21}$ $atoms/cm^{2}$. Fig. \ref{fig:par1}-\ref{fig:par2} show the results: every column of plots illustrates the effect of varying the value of one particular parameter. The number between square brackets denotes the value of the parameter used for all the other computations. \\
In all the plots the shape of soft and hard spectra (dashed and solid line, respectively) are different. This is mainly because of the difference between soft and hard emissivity index ($\gamma_s$ and $\gamma_h$, respectively). \\
The left column of Fig. \ref{fig:par1} shows that when $N_{var} = 0$ (red line), i.e. excluding disc variability and considering only mass accretion rate fluctuations propagating in the hot flow, the model converges to the single hump power spectrum version described in RIK14. Increasing $N_{var}$, the second, lower-frequency, hump starts being distinguishable in all the Fourier products. The disc variability high and low frequency break are the maximum viscous frequency in the disc $\nu_{d,max}$ and $\nu_{v,disc} (r_d)$ respectively (where $r_d = r_o + 5\Delta d$). $\nu_{d,max}$ and $\nu_{v,disc} (r_d)$ are both fixed in this case, so that varying $N_{var}$ does not affect the characteristic frequency of the low frequency hump. The phase lags show a different behavior: the peak frequency correlates with $N_{var}$ and at low frequency ($\nu \approx 0.01$ Hz) the phase lags keep the same profile for all the $N_{var}$ values. The relation between $N_{var}$ and the phase lag profile depends on the function describing the variability in the disc, on the disc temperature profile, and on the selected energy bands. In our case, the amount of variability generated in the disc is described by a Gaussian peaking at the truncation radius, with amplitude proportional to $N_{var}$, and width $\Delta d$ (see Appendix A). When we increase $N_{var}$, the variability increases more steeply with smaller radius close to the truncation radius. \\
If we increase $\Delta d$ (Fig. \ref{fig:par1}, right column) we observe in the power and cross-spectrum that the low frequency hump peak frequency decreases and its profile becomes broader. This is because extending the propagating region towards the edge of the disc, the longer time scale variability contribution becomes more important. For the same reason, increasing $\Delta d$ causes the phase lags to increase at lower frequency. \\
The left column of Fig. \ref{fig:par2} shows the effects of varying the maximum viscous frequency in the disc ($\nu_{d,max}$). Varying $\nu_{d,max}$ affects the size of the viscous frequency trend discontinuity between the disc and the hot flow (i.e. Fig. \ref{fig:nurms}). The bigger this jump the more evident is the two-hump profile. Decreasing $\nu_{d,max}$ causes the low frequency hump peak to move to lower frequency. The phase lag peak frequency correlates with $\nu_{d,max}$, as expected.\\
Finally, the right column of Fig. \ref{fig:par2} shows the effects of varying the maximum temperature in the disc ($T_{d,max}$). Increasing $T_{d,max}$ from 0.05 to 0.1 keV has little consequence on the power and cross spectrum, while we see more evident changes when $T_{d,max}$ varies from 0.1 to 0.5 keV. In particular, the frequency of the phase lag peak is larger than the maximum viscous frequency in the disc ($\nu_{d,max}=0.1$) and it moves to lower frequencies when $T_{d,max}$ increases. \\
This behavior can be explained looking at Fig. \ref{fig:bb}. This plot shows the blackbody emission coming from the ring generating 0.1 Hz variability. When the disc temperature is between 0.05 and 0.1 keV, the blackbody emission is almost entirely detected in the soft band. Because a small fraction of photons is still detected in the hard band, the phase lag at 0.1 Hz is a bit lower than its potential maximum (we expect maximum phase lag when the soft band perfectly matches the disc emission, so when $x_s$ = 1), so that the phase lag peak appears to be at frequencies larger than 0.1 Hz. Increasing $T_{d,max}$ from 0.1 to 0.5 keV means moving the blackbody peak to higher energies, so that a larger part of the blackbody emission is now detected in the hard band. This means that the 0.1 Hz variability is more diluted when $T_{d,max}$ = 0.5-0.7 keV than for lower temperatures and, as a consequence, we observe phase lag suppression at this frequency (black vertical line in the right column of Fig. \ref{fig:par2}). \\

\section[]{Observations and data analysis}
\label{sec:obs}

We analyzed data from the \textit{X-Ray Telescope} (XRT; Burrows et al. 2005) on board of the \textit{Swift} satellite using 5 pointed observations collected between 2010 September 25 and 28 (MJD 55464 - 55467, first 5 observations from the beginning of the outburst). The selected observations contain between $\approx 29.6$ and $\approx 58.0$ ks of data obtained in the Window Timing mode configuration (WT mode), with a time resolution of 1.779 ms. Each observation contains between 1 and 27 Good Time Intervals (GTIs) of $\approx 0.8-2.6$ ks. The data reduction for X-ray spectral analysis was performed using HEASOFT 6.13. The observations were processed using \texttt{xrtpipeline} and the latest \textit{Swift} CALDB files. \\
Source and background spectra were generated in the 0.5-10 keV range; exposure maps and response files were created as outlined in Reynolds \& Miller (2013).  \\
For every GTI we computed soft (0.5 -2.0 keV) and hard (2.0-10 keV) band light curves following the procedure described in Kalamkar et al. (2013): we determined the source and the background region on the CCD and we extracted the light curve for both regions as described in Evans et al. (2007). Each light curve is pile-up corrected. We calculated Leahy-normalized power spectra in the soft and hard band considering 233.19 s data segments in the source light curves, giving a frequency resolution of $\approx 4.3$ mHz and a Nyquist frequency of $\approx 281$ Hz. Using the same segments, we computed Leahy-normalized and source fractional $rms$ normalized (RMS) cross-spectra between soft and hard band in the following way:
\begin{equation}
	\begin{array}{llll}
Leahy: & C_{L}(\nu) & = & \frac{2}{\sqrt{T_s T_h}} F_h (\nu)^{*}  F_s (\nu) \\
RMS: & C_{RMS}(\nu) & = & \frac{\sqrt{T_s T_h}}{(T_s - N_s) (T_h - N_h)} C_{L}(\nu)
	\end{array}
\end{equation}
where $F_{s}(\nu)$ ($F_{h}(\nu)$) represents the Fourier amplitude in the soft (hard) band, and $T_s$ ($T_h$) and $N_s$ ($N_h$) are total and background photons in the soft (hard) band respectively. Using this definition, when $F_s(\nu) = F_h(\nu) = F(\nu)$ the cross-spectrum reduces to the power spectrum with the well known Leahy and RMS normalization (Leahy et al. 1983; van der Klis 1995). For every GTI, Leahy power spectra and cross-spectra were averaged, Poisson noise subtracted estimating the noise level from the power and cross amplitude between 70 and 100 Hz (where no source variability is observed), and finally renormalized to source fractional $rms$ normalization. For every GTI and energy band, we computed total and background count rate ($T$ and $N$) from the source and the background light curve, respectively.\\
We fitted the energy spectra extracted from every GTI in the full energy band with the model \texttt{phabs(diskbb+comptt)} (Mitsuda et al. 1984; Titarchuk 1994). The neutral hydrogen absorption is modeled via \texttt{phabs} with Balucinska-Church \& McCammon (1992) abundances and Asplund et al. (2009) cross sections. The \texttt{comptt} input seed photon temperature is fixed to the disc temperature. The energy spectral analysis and the spectral fit results are those of Kalamkar et al. (2015b). From energy spectra fitting of every GTI we computed the fraction of disc photons emitted in the soft band ($x_s$) and we obtained the maximum disc temperature in the disc ($T_{d,max}$). \\

\section{Results}
\label{sec:res}

We fit our model simultaneously to the soft and hard band (0.5-2 keV and 2-10 keV, respectively) power spectrum, and the cross-spectrum between these two energy bands, for the 5 pointed \textit{Swift} observations described in the previous section. We perform fits to the real and imaginary parts of the cross-spectrum, since this has statistically favorable properties, but plot in terms of amplitude and phase, which is more intuitive. These observations display enhanced low frequency variability in their power spectra, which Kalamkar et al. (2015a) suggested may result from disc variability. However, MAXI J1659-152 displays absorption dips which may influence the low frequency variability properties (Kuulkers et al. 2013). In this section, we first investigate these absorption dips before using \textsc{propfluc} to determine if the low frequency variability in these observations can originate from propagating fluctuations in the disc. 

\subsection{Dip and no-dip regions}
\label{sec:dnd}

\begin{figure}
\center
\includegraphics[scale=0.4, angle=180,trim={2cm 2.cm 1.cm 5cm},clip]{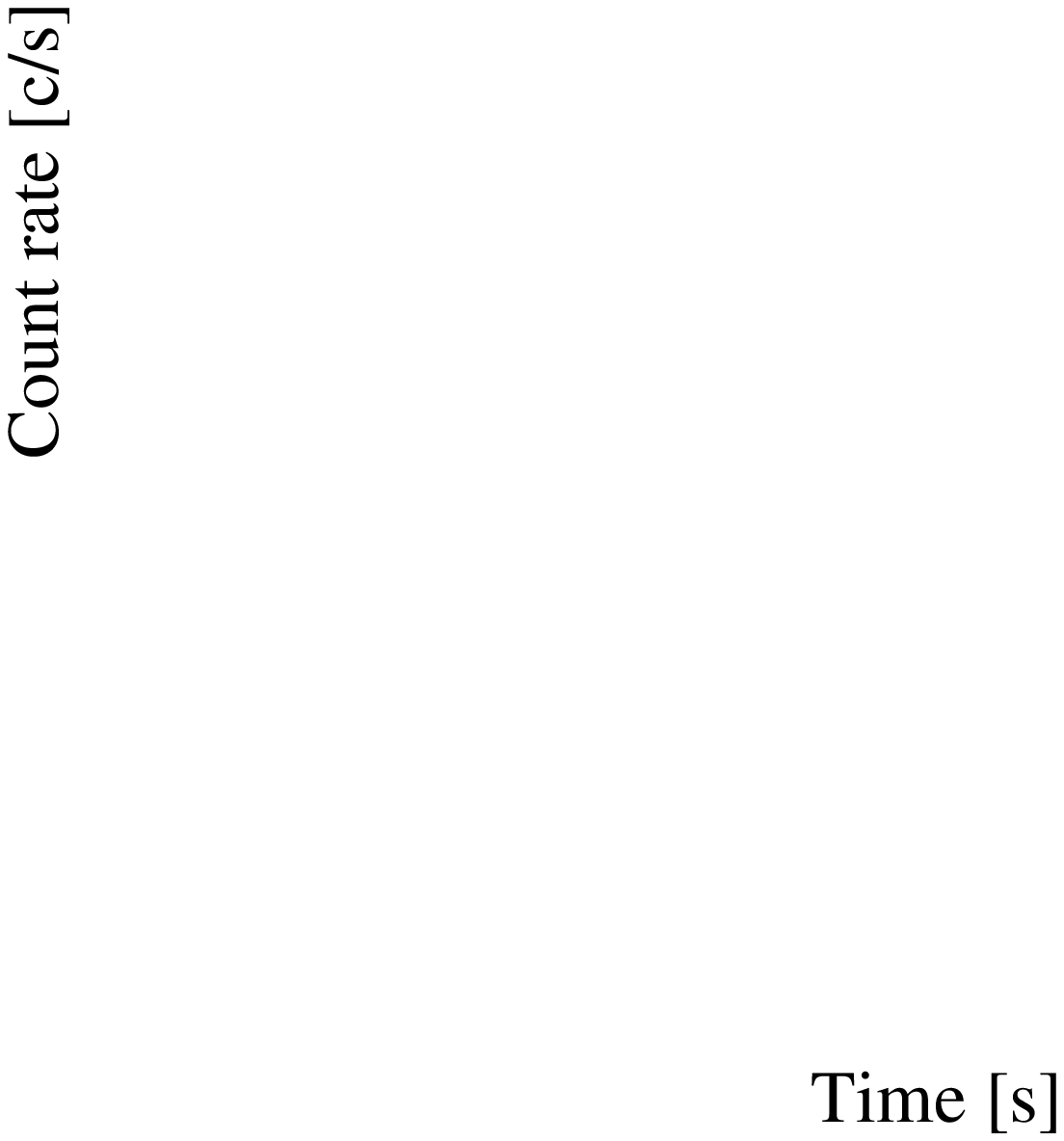}
\caption{Light-curve of the fourth observation from the beginning of the outburst. The light curve consists of 9 GTIs (clusters of points inside ellipses), where every point corresponds to count rate averaged over 233.19 s (the same time interval used for performing Fourier analysis). Several GTIs show clear dipping behavior (blue dashed ellipses). In our analysis we considered only no-dip time regions (red ellipses).}
\label{fig:obs3flux}
\end{figure}

\begin{figure} 
\center
\includegraphics[scale=0.75, angle=-90,trim={0cm 9.5cm 5cm 0cm},clip]{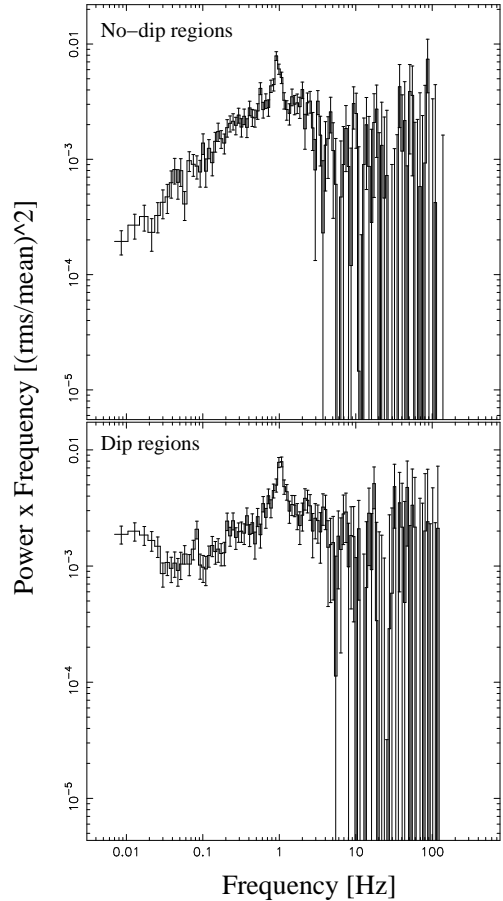}
\caption{Power spectra computed on 233.19 s time intervals from the fourth observation from the beginning of the outburst. The power spectrum of GTIs showing clear dipping behavior (Dip regions) shows extra low frequency variability compared to no-dip regions.}
\label{fig:dippsa}
\end{figure}
The MAXI J1659-152 2011 outburst light curve shows two types of peculiar intensity variations: absorption dips and transition dips (Kuulkers et al. 2013). The first kind of dips, the absorption dips, is observed at day 0.3 up to day 8.2 from the beginning of the outburst (MJD 55464), the second kind of dips, the transition (Kuulkers et al. 2013) or "flip-flop" (Kalamkar et al. 2011) dips, are observed sporadically from day 23.7, so they are not included in the time period we analyzed. The depth of the absorption dips is between about 50\% and 90\% the average out-of-dip interval intensity. During these dips, the source hardens, and the deeper the dip the stronger the hardening. The dips become shallower as the source intensity increases along the outburst. The dip occurrence can be explained with the presence of some absorber in the disc that periodically obscures emission along the line of sight. This period is associated with the orbital period of the binary, which allowed Kuulkers et al. (2013) to estimate with high precision the period of the system ($2.414 \pm 0.005$ hrs) and to estimate lower and upper limit values for the disc inclination ($\approx 65^{\circ}$ and $80^{\circ}$, respectively).\\
\begin{figure} 
\center
\includegraphics[scale=0.5,angle=-90, trim={0cm 0cm 0 0cm},clip]{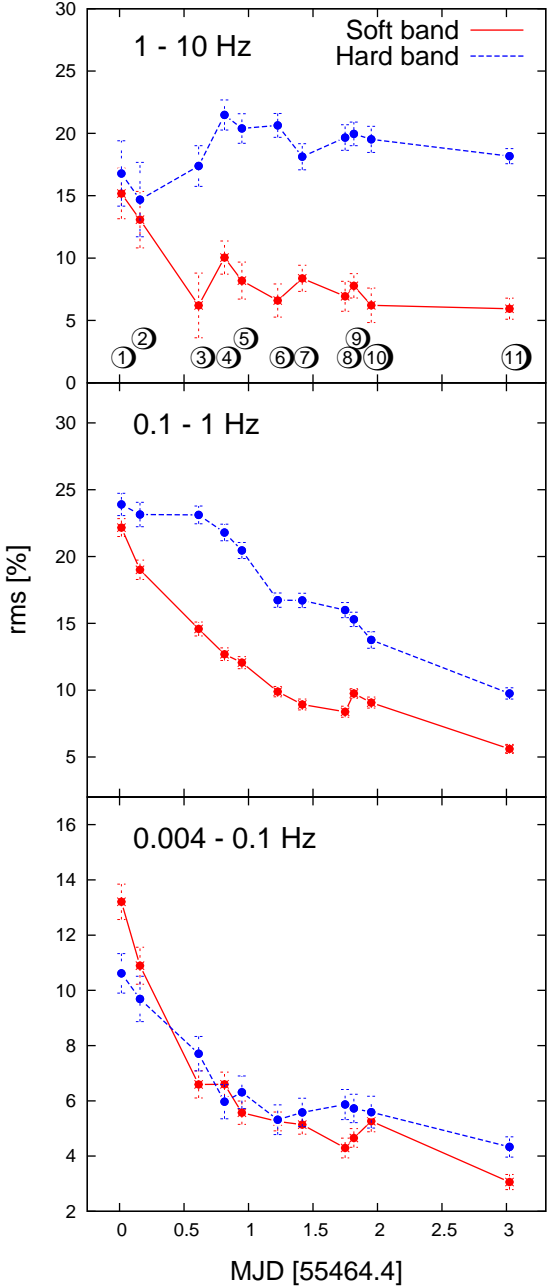}
\caption{Fractional $rms$ amplitude computed from power spectra in three different frequency bands and in two energy bands. Every point in the plots corresponds to a single $no-dip$ time interval of the source light curve.}
\label{fig:rmsspec}
\end{figure}
The XRT observations of MAXI J1659-152 we analyzed clearly show this dipping behavior. Fig. \ref{fig:obs3flux} shows the light curve of the fourth observation from the beginning of the outburst (ObsID: 00434928003, starting at MJD 55466), each point in the plot represents the count rate in the full band (0.5 - 10.0 keV) averaged over a 233.19 s time interval (the same segment we used to compute timing analysis products). This observation consists of 9 GTIs (blue and red ellipses in the plot) and some of them are characterized by an intensity drop (blue ellipses). The light curve of the source during the dips is characterized by strong $\sim$ 100 s time scale variability, so that averaged power spectra of GTIs including dips show extra low frequency noise between $\sim$ 0.01 and $\sim$ 0.1 Hz (see Fig. \ref{fig:dippsa}). This extra low frequency variability may be due to fluctuations in the absorbing material. In any case, it is not intrinsic to the accretion flow, since it is only present during the absorption dips. For this reason, we excluded GTIs including dips from every observation, leaving a deeper analysis on comparison between spectral and timing properties of dip and no-dip regions to future work.\\
The power spectra of MAXI J1659-152 in the no-dip regions are still characterized by different broad power spectral components: a low frequency component between $\sim$ 0.1 and 1 Hz and a main hump with characteristic frequency between $\sim$ 1 and 5 Hz (power spectral components referred as ``break" and ``hump" in Kalamkar et al. 2015a, respectively). This low frequency hump in the out-of-dip power spectrum may be driven by disc variability, with the high frequency hump generated in the flow. Before testing this hypothesis in the next section, we first consider if, instead, even the $\sim$ 0.1-1 Hz hump results from residual dipping activity that we have not been able to "weed out" with our GTI selections. If this were the case, the soft band would be more variable than the hard band (as during the dips), as changes in the column density of the absorbing material affect predominantly the soft X-rays. Fig. \ref{fig:rmsspec} shows the fractional $rms$ amplitude computed in three different frequency bands for every single \textit{no-dip} GTI selected in our analysis in the soft and hard band. Looking at the variability amplitude between 0.1 and 10 Hz (Fig. \ref{fig:rmsspec} top and middle panel), we notice that it is always larger in the hard band. This excludes absorption mechanisms as the origin of the ``break" and ``hump" component.\\
We notice that the 1$^{st}$ and 2$^{nd}$ no-dip GTIs show extra low frequency ($<$ 0.1 Hz) variability (first two points in Fig. \ref{fig:rmsspec}, bottom panel). Because this variability is larger in the soft band, it is still possibly due to some residual absorption dip in the selected GTI. The presence of extra low frequency variability due to the dips can influence our test on the hypothesis of mass accretion rate fluctuations coming from the disc, for this reason any consideration regarding these two GTIs has to be handled with care.\\

\subsection{A soft low-frequency QPO detected in GTI1}
\begin{figure}
\center
\includegraphics[scale=0.4, angle=-90,trim=1cm 4cm 1cm 3cm,clip]{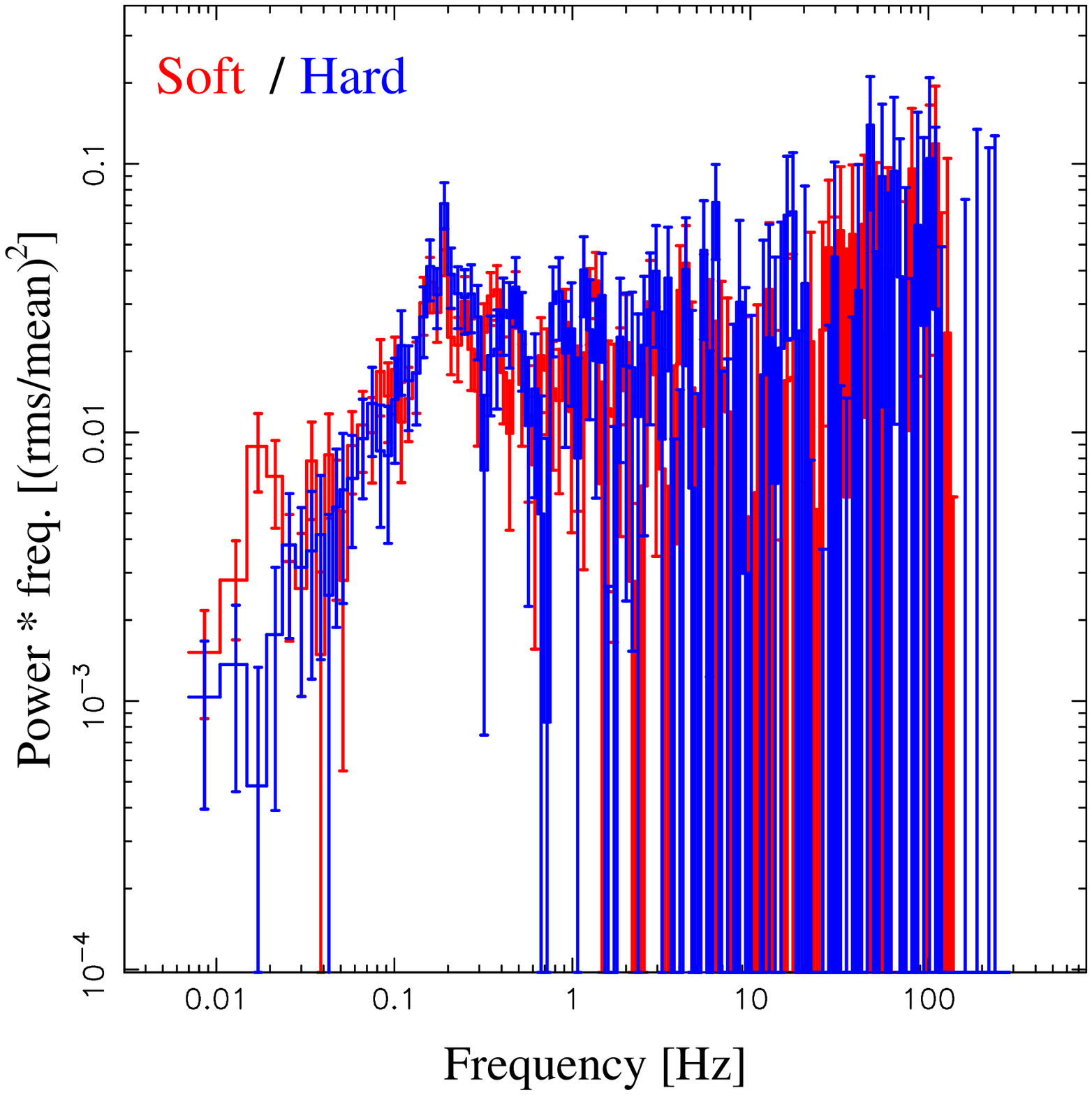}
\caption{Soft (red) and hard (blue) power spectrum of the first GTI. This GTI is characterized by extra low frequency variability in the soft band, the reason of this extra variability is the presence of a soft low-frequency QPO at 0.018 Hz.}
\label{fig:softqpo}
\end{figure}

We report the detection of a significant QPO (4.22$\sigma$) in the soft power spectrum of the first observation (GTI1, see Fig. \ref{fig:softqpo}). We obtained the QPO characteristics by multi-Lorentzian fit (Belloni, Psaltis \& van der Klis 2002): $rms = 7.33 \pm 0.80 \%$, coherence $Q = 2.27 \pm 0.92$, and $\nu_{max} = 0.018 \pm 0.001 Hz$. The feature is similar to the 11 mHz QPO observed in two RXTE observations (2-60 keV) of the BH candidate H1743-322 by Altamirano \& Strohmayer (2012).

\subsection{\textsc{propfluc} fits}
\label{sec:modfit}

\begin{figure*} 
\center
\includegraphics[scale=0.64, angle=-90,trim=1cm 1.2cm 1cm 0cm,clip]{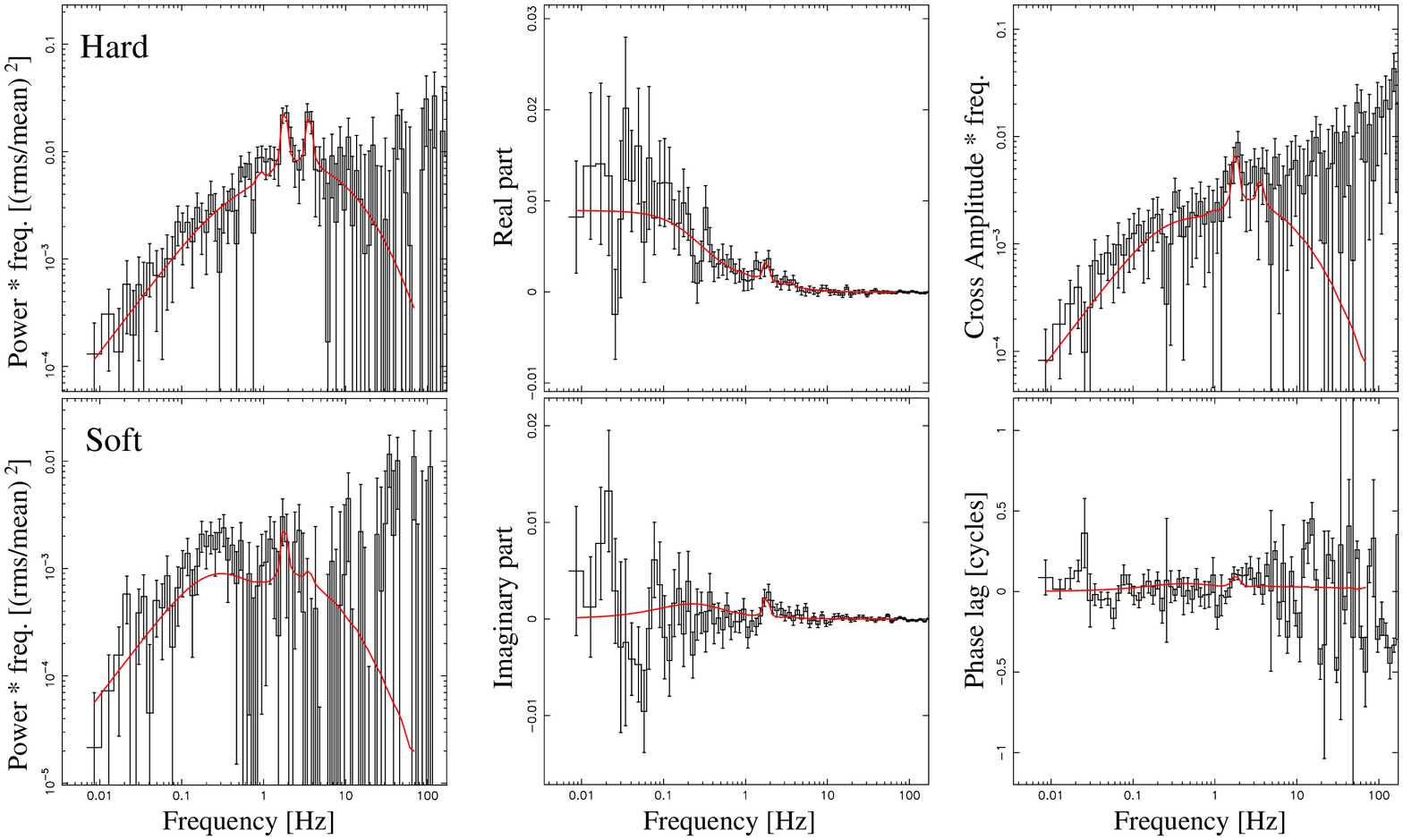}
\caption{Example of  joint fit of the 5$^{th}$ observation (GTI11). Fits are performed on the hard and soft power spectrum (1$^{st}$ column), and on the real and imaginary part of the cross-spectrum (2$^{nd}$ column). The 3$^{rd}$ column shows the cross-spectrum represented in terms of amplitude and the phase lag, rather than real and imaginary parts.}
\label{fig:fitex}
\end{figure*}

\begin{figure} 
\center
\includegraphics[scale=0.4, angle=270,trim=0cm 0.cm 0cm 0cm,clip]{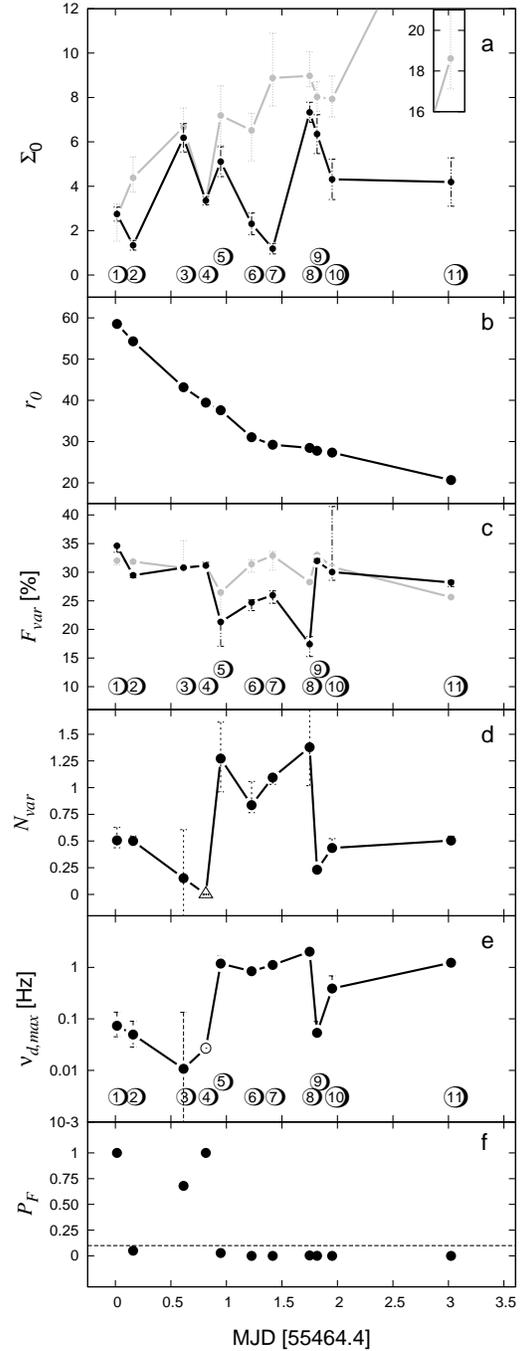}
\caption{\textsc{propfluc} best fit parameters versus time (black points). All the points were plotted with 1$\sigma$ error bars. Grey points correspond to single hump best fit parameters (fit results excluding the disc propagating region). The filtered GTIs are indicated with integers from 1 to 11 (symbols in panels $a$, $c$, and $e$). Panel $f$ indicates the F probability related to a single hump fit.}
\label{fig:fit}
\end{figure}

We fitted logarithmically binned data points in the frequency range 0.004-70 Hz, using the same resolution for data and model. For every observation we fitted simultaneously the soft and hard spectrum, and the cross-spectrum between the two energy bands using $N_{dec} = 35$ for all the fits. A minimum of 30 rings is required to avoid interference patterns at high frequency (IK13) and we confirmed experimentally that a higher radial resolution did not produce any significant difference in $\chi^2_{red}$. We combined the QPO with the broad band variability by addition (instead of multiplication, see IK13 and Appendix \ref{qpo}). For all the fits we fixed the surface density profile in the hot flow (the parameters $\zeta$, $\gamma$, and $\lambda$), the transition radius of the smoothly broken power law $r_{bw}$, the emissivity in the soft and hard band ($\gamma_s$, $\gamma_h$), the mass M, and the dimensionless spin parameter of the BH $a_*$. In our analysis we used the \textit{Swift} Redistribution Matrix File and Ancillary Response File closest to our data and we considered a column density of $n_H = 1.7 \times 10^{21}$ atoms/cm$^2$ (Kalberla et al. 2005). For fitting the low frequency hump we fixed $\Delta d$ to 35 corresponding to $r_d \approx 235$ assuming $r_o \approx 60$ (the largest $r_o$ value fitted in the observations we analyzed).  In Sec. \ref{sec:exp} we showed how the radial extent of the disc affects both the frequency and the integrated power of the low frequency hump, so that in general $\nu_{d,max}$ and $N_{var}$ variations can be interpreted involving $\Delta d$ changes. The choice of fixing $\Delta d$ can be justified considering that larger $\Delta d$ values would produce a broader low frequency component than the one observed in the data and $\nu_{d,max}$ values closer to or even larger than the viscous frequency in the hot flow at the truncation radius $\nu_{v, flow} (r_o)$. This last configuration is not consistent with the double hump model assumptions which: 1) imply a discontinuity in the physical properties of the entire accreting region at the truncation radius (so a ``jump" of the viscous frequency at $r_o$), and 2) imply that the characteristic time scale of the variability originating in the hot flow is shorter than in the disc. \\
We computed the hardness ratio $HR$ dividing hard by soft background subtracted photon counts, and the disc fraction in the soft band $x_s$ from spectral fitting. The free fit parameters are the surface density normalization constant $\Sigma_0$, the fractional variability in the hot flow $F_{var}$, the truncation radius $r_o$, the $rms$ and phase lag of the main QPO, second, and eventually third and sub-harmonic ($\sigma_{qpo}$, $\sigma_{qpo2}$, $\sigma_{qpo3}$, $\sigma_{sub}$ and $\phi_{qpo}$, $\phi_{qpo2}$, $\phi_{qpo3}$, $\phi_{sub}$ respectively) in both soft and hard band, the variability in the disc as a fraction of the hot flow variability $N_{var}$, and the maximum viscous frequency in the disc $\nu_{d,max}$.\\
Fig. \ref{fig:fitex} shows the simultaneous \textsc{propfluc} fit of the soft and hard power, and the cross-spectrum of the fifth observation (in particular, we plot the hard and soft power spectrum, the real and imaginary part of the cross-spectrum, the cross-spectrum, and the phase lag). The best fit parameters are reported in Tab. \ref{tab:res}, and Fig. \ref{fig:fit} shows the evolution of the free model parameters with time (black points). For simplicity, we labeled with integers from 1 to 11 the no-dip GTIs we filtered from the 5 selected pointed \textit{Swift} observations (Fig. \ref{fig:fit} panel $a$, $c$, and $e$). From GTI 1 to 11, $\Sigma_0$ increases from $\approx$ 2.8 to $\approx$ 4.2. This increasing trend is not continuous, there are 3 dips (values smaller than the contiguous observations) at GTI 2, 4 and 7, and 3 local peaks at GTI 3, 5, and 8. The truncation radius $r_o$ shows a clear smooth decreasing trend from $\approx$ 60 to 20, indicating an average truncation radius recession speed of about 0.6 $R_g $/h ($\approx$ 9 km/h). The fractional variability $F_{var}$ shows a smooth decreasing trend between GTI 1 and 11 (from $\approx$ 35\% to 28\%) interrupted by values smaller than average ($\approx$ 17-26\%) between GTI 5 and 8. Panels $d$ and $e$ in Fig. \ref{fig:fit} show the evolution of the model parameters related to the low frequency hump, empty symbols represent upper limits ($3 \sigma$ confidence level). $N_{var}$ varies around $\approx$ 0.2 and 0.5 in GTI 1-4 and 9-11, respectively. Between GTI 5 and 8, $N_{var}$ is characterized by larger values ($\approx$ 0.8-1.3), in particular it shows an increasing trend between GTI 6 and 8. The maximum viscous frequency in the disc $\nu_{d,max}$ shows an evolution similar to $N_{var}$  with a general increasing trend  (from $\approx$ 0.07 to 1.23 Hz) and higher values than average between GTI 5 and 8. 

\section{Discussion}
\label{sec:dis}

\subsection{Double hump and single hump power spectrum} 

\begin{figure} 
\center
\includegraphics[scale=0.5,angle=270]{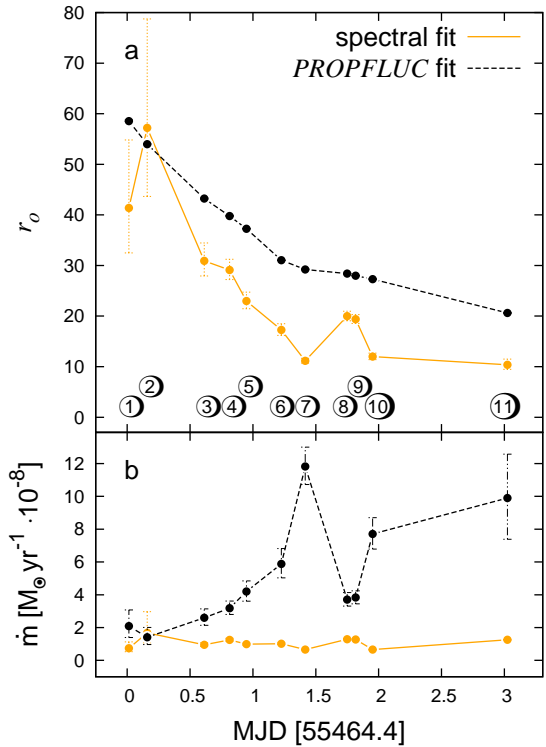}
\caption{Truncation radius $r_o$ ($a$) and mass accretion rate ($b$) computed using timing (black dashed line) and spectral fit (yellow solid line) results.}
\label{fig:mdot}
\end{figure}
In this study, we presented a new version of the \textsc{propfluc} model that can produce a two-hump power spectrum, where the main hump originates because of mass accretion rate fluctuations propagating through the hot flow (as described in ID12, IK13, RIK14) and an additional low frequency hump is produced by fluctuations propagating from the thermal varying disc into the hot flow. Assuming a photon emission mechanism for the hot flow and the disc, we calculated power spectra and cross-spectra between two energy bands (Sec. \ref{sec:mod}). We used this model to study 5 observations of the BH MAXI J1659-152 during its 2010 outburst using \textit{Swift} data. We measured the spectral parameters required as input by the model by spectral fits, and we fitted soft and hard band power spectra, and cross-spectra simultaneously with both the single and the double hump \textsc{propfluc} version. In a single hump power spectrum, mass accretion rate fluctuations generated and propagating in the hot flow are the only variability source. In a double hump power spectrum, variability is generated both in the hot flow and in the disc, so that the way the total variability power is distributed between low frequency and main hump depends both on the hot flow and disc characteristics. If a low frequency component is present in the power spectrum, we expect the peak of the main hump in the two-hump model to be shifted to higher frequency relative to the single hump model (i.e., when the low frequency component is not taken into account). The grey points in Fig. \ref{fig:fit}$a$ and $c$ are the best fit parameter values obtained using a single hump power spectrum. We find that both $\Sigma_0$ and $F_{var}$ from single hump fit are different compared to the double hump fit results, in particular the single hump $\Sigma_0$ values are larger for all the GTIs. \\
Fig. \ref{fig:fit}$f$ shows the F probability for every pair of fits, i.e., the probability that the $\chi^2$ improvement when using the double hump model is due to statistical fluctuations. Low probability indicates that the double hump model gives a better fit than the single hump model. $P_F$ exceeds 10\% only in 3 cases (GTI 1, 3, and 4), for all the other GTIs the use of the additional hump originating in the disc is statistically justified. \\
So, the double hump model fits significantly better than the single hump one, but the fits are not formally acceptable (see Tab. \ref{tab:res} at reduced $\chi^2$ values of $\approx$ 1.3 - 1.5). Given that in this first-ever attempt to quantitatively model power and cross-spectra jointly our model reproduces most of the overall characteristics of the data, it seems useful to discuss the evolution of the source in terms of the model physical parameters, but we note that any interpretation of the parameter evolution has to be handled with care.\\

\subsection{Evolution of the physical parameters}
Contrary to the case with previous applications of the model (ID12, RIK14), the fractional variability shows in general a decreasing trend. \\
\begin{figure} 
\center
\includegraphics[scale=0.5, angle=270,trim=0cm 0.cm 0cm 0cm,clip]{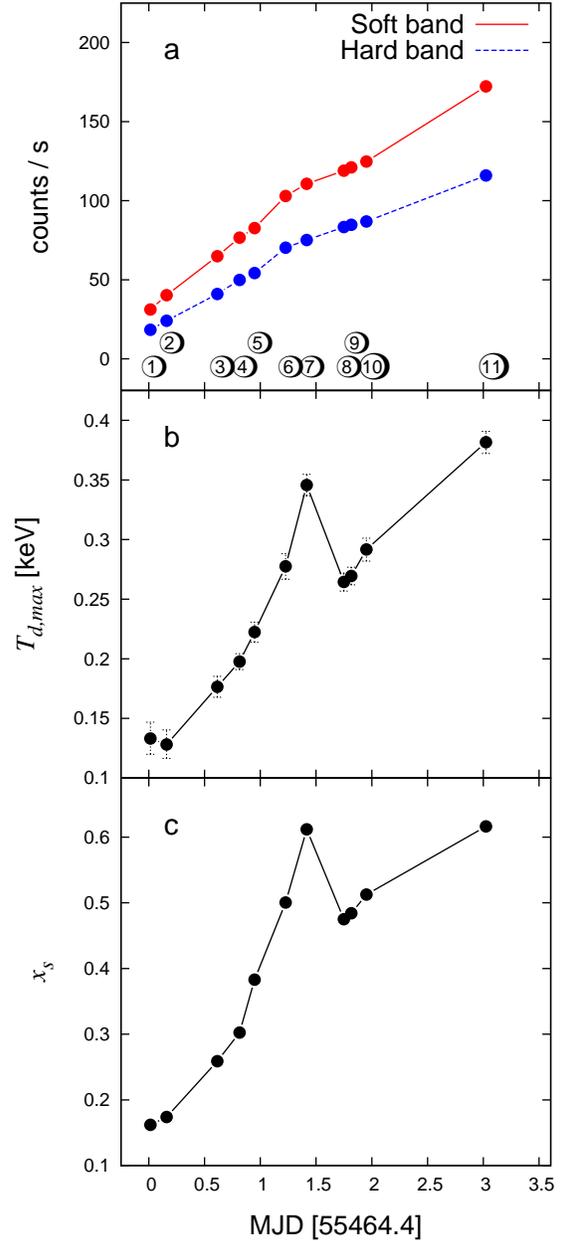}
\caption{Spectral properties of the source for all the analyzed observations. a) Count rate in the soft and hard energy band; b) Maximum disc temperature in the disc; c) Fraction of total photons detected in the soft band.}
\label{fig:spec}
\end{figure}
For a fixed ring, $\Sigma_0$ is proportional to the surface density profile divided by mass accretion rate (RIK14), so that, assuming a constant surface density per ring, $\Sigma_0$ can trace variations in mass accretion rate. Even if $\Sigma_0$ shows a general increasing trend from GTI 1 to 11, the several dips and peaks in its trend suggest a variable accretion regime, in particular between GTI 7 and 10.   \\ 
We did not detect a significant lag associated with the low frequency component. This may seem surprising if the process generating the broad band noise is mass accretion rate fluctuations propagating through the accreting flow. In a propagating fluctuations model the amplitude of the lag between two energy bands depends on the radial extension of the propagating region, on the difference between the emissivity profiles, and on the propagation time scale of the fluctuations (Nowak et al. 1999b; Kotov et al. 2001; Arevalo \& Uttley 2006). In our model, the propagation time is set by the local viscous time scale, that is equal to the characteristic time scale of the variability produced in every ring. So, the main \textsc{propfluc} parameters affecting the amplitude of the phase lag are $\gamma_s$ and $\gamma_h$ (fixed in our fit), and $T_{d,max}$, $x_s$, and $HR$ (estimated by spectral fitting and measuring the photon counts in the soft and hard bands, see Fig. \ref{fig:spec}). In Sec. \ref{sec:exp} we showed that for $x_s$ = 0.9 and $HR$ = 1, \textsc{propfluc} predicts $\approx$ 0.1 cycle phase lags in the low frequency hump. In all our observations $x_s < 0.63$ and $HR < 0.7$, leading to predicted phase lags smaller than 0.1 cycles. Such lags are not detectable in our data. So, the absence of phase lags in the low frequency hump does not exclude the hypothesis of propagating mass accretion rate fluctuations and it is consistent with \textsc{propfluc} predictions for this case. However, because of this lack of additional information, we can not completely remove the degeneracy between the model parameters, especially those related to the radial extension of low frequency hump and affecting the lag profile ($\Delta d$, $N_{var}$, $\nu_{d,max}$).\\

\subsection{Mass accretion rate from spectral and timing analysis}
\begin{figure} 
\center
\includegraphics[scale=0.5,angle=270]{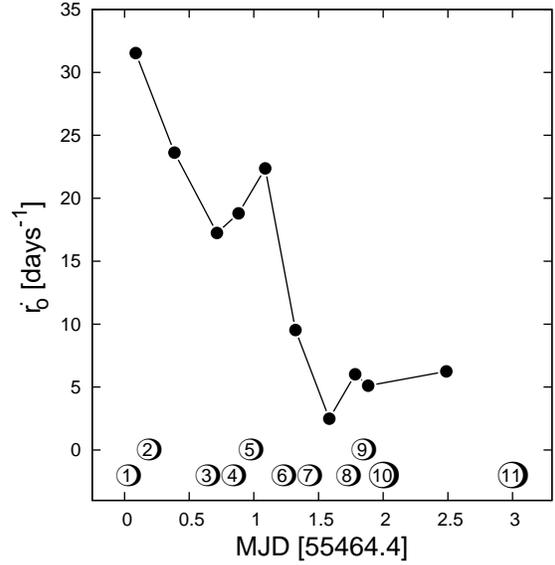}
\caption{The black points represent the average $r_o$ decreasing rate between couples of GTIs (indicated with symbols in the same way of previous Fig.s)}
\label{fig:rate}
\end{figure}

It is possible to compute two independent estimates of the truncation radius $r_o$ from timing and spectral analysis, respectively. In the \textsc{propfluc} fits $r_o$ mainly depends on the QPO frequency; we can also compute $r_o$ from spectral fits using the \textsc{diskbb} model normalization , the distance $d$, and the inclination of the accreting disc to the line of sight $\theta$ ($norm = \big( \frac{r_o/[km]}{d/[10kpc]} \big)^2 cos(\theta)$). Fig. \ref{fig:mdot}$a$ shows the truncation radius computed using both \textsc{propfluc} (black dashed line) and spectral fit (yellow solid line). For computing the last one we used $d$ = 8.6 $\pm$ 3.7 kpc and $\theta$ = $\ang{72.5}$ $\pm$ $\ang{7.5}$ (Kuulkers et al. 2013). The absolute values shown in the plot have to be handled with care: the QPO frequency depends not only on $r_o$, but also on the hot flow structure ($r_{bw}$, $\lambda$, $\zeta$, $\kappa$), on the hot flow inner boundary $r_i$, on the spin, and on the mass of the BH; the estimation of $r_o$ via the \textsc{diskbb} normalization needs to be corrected because of possible deviation of the emitting region from pure blackbody behavior (Merloni, Fabian, \& Ross 2000), and general relativistic effects (e.g. Ebisawa et al. 1994). However, $r_o$ values obtained in these two ways are the same within a factor of 2, and, in general, show the same decreasing trend. The main difference between the two is that $r_o$ from \textsc{propfluc} fit is characterized by a smooth decreasing trend, while $r_o$ from spectral fit shows a more irregular behavior with a local peak in GTI 8-9.\\
From the truncation radius and the maximum temperature in the disc, it is possible to compute the mass accretion rate $\dot{M}$ (Frank, King, \& Raine 2002):
 \begin{equation}
\dot{M} = \frac{8 \pi \sigma}{3 G M_{BH}} r_o^3 T_{max}^4
\label{eqn:mdot}
\end{equation}
Similarly to Fig. \ref{fig:mdot}$a$, Fig. \ref{fig:mdot}$b$ shows $\dot{M}$ computed from the \textsc{propfluc} $r_o$ (hereafter $\dot{M}_{prop}$) and the spectral fit $r_o$ measurement (hereafter $\dot{M}_{spec}$). $\dot{M}_{spec}$ is almost constant while $\dot{M}_{prop}$ shows an increasing trend with a peak at GTI 7 followed by a dip in GTIs 8-9. This difference is because $T_{d,max}$ and \textsc{diskbb} normalization variations compensate each other, while the decreasing $r_o$ trend obtained from timing analysis is smooth, so that the resulting $\dot{M}_{prop}$ varies similarly to $T_{d,max}$ (Fig. \ref{fig:spec}$b$). Looking at luminosity variations (Fig. \ref{fig:spec}$a$), $\dot{M}_{prop}$ is in agreement with the increasing trend in count rate. It is also interesting to compare the jump followed by the dip in $\dot{M}_{prop}$ between GTIs 6 and 10 with similar features over the same GTIs in $\Sigma_0$, $F_{var}$, $N_{var}$, and $\nu_{d,max}$ (Fig. \ref{fig:fit}; we note that these are \textsc{propfluc} parameters independent from spectral fit parameters). The peak and the following dip in mass accretion rate rate at GTI 7 and 8-9 respectively are consistent with $\Sigma_0$ variations: higher $\dot{M}_{prop}$ values correspond to lower $\Sigma_0$ values, and vice versa. The relation between $\dot{M}_{prop}$ and $F_{var}$, $N_{var}$, and $\nu_{d,max}$ is more difficult to interpret. We may speculate that the jump in $\dot{M}$ at GTI 7 (triggered by some instability) may have stirred up extra variability in the disc and in the flow (increasing $N_{var}$ and $F_{var}$) and will have increased the viscous frequency in the disc (since $\nu_{v} \propto \dot{M}$ though mass conservation). The dip in $\dot{M}_{prop}$ (GTI 8) then follows as the supply of material is depleted, leading to the corresponding dips seen in $F_{var}$ (GTI 8), and in $N_{var}$ and $\nu_{d,max}$ with a GTI of delay (the material closer to the BH is depleted faster). Because of the jump in $\dot{M}$ we would also expect some variation in the average decreasing rate of $r_o$. Indeed, in a truncated disc geometry the radial dimension of the hot flow depends on mass accretion rate, and when mass accretion rate increases (rising part of the outburst), $r_o$ moves in. Fig. \ref{fig:rate} shows the rate of $r_o$ decrease for all the observations considered in our analysis. Between GTI 5 and 6 $r_o$ decreases faster than average, while between GTI 7 and 8 it decreases slower than average. This behavior is consistent with the mass accretion rate variations described above.

\subsection{The absorption dips}
In our study of MAXI J1659-152 we excluded time intervals characterized by absorption dips (see Sec. \ref{sec:dnd}). Power spectra computed including dip-regions show strong low frequency noise ($\sim$ 0.01 Hz). This additional noise component was identified as ``lfn" in the Kalamkar et al. (2015a) timing analysis of the source. In their study, it was proposed that the characteristics of the $< 0.1$ Hz ``lfn" component during the outburst could be explained with variability arising in the disc and propagating into the hot flow. Here, we exclude this possibility for the observations we analyzed mainly because the ``lfn" component is strongly coupled with the periodic absorption dips in the light curve (Sec. \ref{sec:dnd}). Contrary to the case of the ``lfn'', the $rms$ amplitude in the 0.1 - 10 Hz frequency band (Fig. \ref{fig:rmsspec}, top and middle panel) is always larger in the hard band. For this reason the components identified as ``break" and ``hump" in Kalamkar et al. (2015a) (with characteristic frequency between $\sim$ 0.1 and $\sim$ 5 Hz in our observation sample), can not have been produced by varying absorption. In our fit with \textsc{propfluc}, the ``break" and ``hump" components are associated to the disc and the hot flow, respectively.\\

\section{Conclusions}

We applied the double hump model \textsc{propfluc} to investigate the HIMS of MAXI J1659-152 using \textit{Swift} data. In the model, low frequency broad band components are interpreted as the result of mass accretion rate fluctuations arising in the disc and propagating towards the BH through the hot precessing flow. This double hump model was statistically preferred to a single hump model for most of the GTIs analyzed. In our analysis we detected only small phase lag associated with the low frequency variability, however model predictions are consistent with the data. Using both spectral and timing analysis we estimated recessing trend in truncation radius, and from that we inferred the mass accretion rate. Considering the truncation radius estimate from the \textsc{propfluc} fit and the maximum temperature in the disc (spectral fit parameter), we found a peak in the average increasing mass accretion rate trend that matches the variability properties of the accreting system (the amount of variability generated and the viscous frequency in the disc and in the flow). Considering the truncation radius estimate from spectral fit, would have lead to an almost constant mass accretion rate, in contrast to observations (the total counts increase almost linearly in our observation sample). Our analysis constitutes the first joint fitting of compact object cross-spectra and power spectra with a single self-consistent physical model. \\
\\
\textbf{\textit{ACKNOLEDGEMENTS}}\\
We thank the anonymous referee for his/her useful comments that greatly helped to improve the manuscript. S. Rapisarda, A. Ingram, and M. van der Klis acknowledge support from the Netherlands Organization for Scientific Research (NWO). M. Kalamkar acknowledges support by Marie Curie FP7-Reintegration-Grant under contract no. 2012-322259. This research has made use of the XRT Data Analysis Software (XRTDAS) developed under the responsibility of the ASI Science Data Center (ASDC), Italy.

\begin{appendix}

\section{new \textsc{propfluc} formalism}
\label{app}

In the hot flow the viscous frequency is described by a smoothly broken power-law, in the disc we assume the following profile (Shakura \& Sunyaev 1973):
\begin{equation}
\label{eqn:nudisk}
\nu_{v,disc} (r) = \nu_{d, max} \left ( \frac{r}{r_o} \right ) ^{-3/2}
\end{equation}
where $\nu_{d, max}$, the maximum viscous frequency in the disc, is a model parameter. \\
We assume the amount of variability generated per radial decade in the hot flow to be constant (ID11, ID12, RIK14), while in the disc it follows a Gaussian profile peaking at the truncation radius $r_o$:
\begin{equation}
\label{eqn:rmsdisk}
\sigma(r) = \sigma_{o} N_{var} e^{-\frac{1}{2} \left ( \frac{r-r_o}{\Delta d} \right )^2)}
\end{equation}
where the model parameters $N_{var}$ and $\Delta d$ govern amplitude of disc variability and radial extension of the varying disc respectively. In particular, considering that $\sigma_0 = F_{var} / \sqrt{N_{dec}}$, $N_{var}$ estimates the amount of variability generated in the disc per radial decade as a fraction of the flow fractional variability $F_{var}$. When $N_{var} = 1$, the variability produced in the innermost ring of the disc is equal to the variability produced in the outermost ring of the hot flow. If the variability produced in the hot flow is larger than in the disc, it follows that $N_{var} < 1$. \\
As described in IK13, the flux observed in some energy band can be written in the following way:
\begin{equation}
\label{eqn:flux}
f_h (t) = \sum_{j=1}^N h(r_j) \dot{m}(r_j,t)
\end{equation}
where the subscript $h$ indicates the hard band (so that $f_h (t)$ is the flux observed in the hard band), $\dot{m}(r_j,t)$ represents the mass accretion rate fluctuations in the ring $r_j$, $h(r_j)$ is the number of photons coming from the ring $r_j$ and detected in the hard band, and $N$ is the total number of rings (disc plus hot flow rings). From the flux, we can compute the power spectrum in the hard band:
\begin{equation}
\label{eqn:ps}
P_h(\nu) = | F_h (\nu) |^2 = \sum^{N}_{i,j=1} h(r_i) h(r_j) \dot{M}^{*}(r_i , \nu)  \dot{M}(r_j , \nu)
\end{equation}
here we adopt the convention that lower case and capital letters indicate time series and Fourier transforms respectively (unless specified differently), and the symbol $^*$ stands for complex conjugation.   \\
The weighting factors $h(r_j)$ depend on how many photons are detected in a selected energy band, so on photons emitted and on the instrument response. The general expression for $h(r_j)$ factors is the following:
\begin{eqnarray}
\begin{aligned}
h(r) & \propto  \sum^{I_{max}}_{I_{min}} Q(r , I) \\
       & =  \int_0^{\infty}A(r, E)  M_a (E) dE  \sum^{I_{max}}_{I_{min}} R_D (I , E) 
\end{aligned}
\label{eqn:hrs}
\end{eqnarray}
where $Q(r, I)$ represents the photon counts measured from ring $r$ by our detector in the $I^{th}$ energy channel, $I_{min}$ and $I_{max}$ are the boundaries of the channel range selected, $A (r , E)$ is the spectrum emitted from the radial bin centered at $r$ (photons per unit time, unit energy, and unit telescope collecting area), $M_a (E)$ is the dimensionless absorption model, and $R_D (I , E)$ is the response of the detecting instrument. \\
In the new version of $\textsc{propfluc}$, we assume that the spectrum emitted from a certain radius $r$ in the disc is a blackbody with temperature $T(r) = T_{max} (r/r_o)^{-3/4}$ (Shakura \& Sunyaev 1973). With this assumption, $A(r, E)$ is given by the Planck function. \\
In the absence of a full model for the spectrum emitted by the flow, we parametrize the weighting factors in the following way (IK13):
\begin{equation}
\label{eqn:hrf}
h_{flow} (r)  \propto r^{(2-\gamma_h)} \Sigma(r) 
\end{equation}
where $\gamma_h$ is the emissivity index for the hard energy band, and $\Sigma(r)$ is the surface density in the flow. The $h(r)$ values described so far have an arbitrary normalization, but in order to compare \textsc{propfluc} computations with real data, we need to include some other parameter taking into account the spectral characteristics of the source. Considering two energy bands, soft and hard ($s$ and $h$ respectively), we define the disc fraction in the soft band $x_s$ as the fraction of total photons in the soft band that comes from the disc. $x_s$ is a model parameter and can be estimated from spectral fitting. From $x_s$ and the hardness ratio $HR$ (the ratio between hard and soft photon counts), it is possible to compute the disc fraction in the hard band.
\begin{equation}
x_h = \frac{h_d / s_d} {HR} x_s
\end{equation}
where $s_d \equiv \sum^{N_{disc}}_{j=N_{flow}} s(r_j)$, $h_d \equiv \sum^{N_{disc}}_{j=N_{flow}} h(r_j)$. Using the disc fraction in the hard band it is possible to normalize the $h(r)$ values of Eq. \ref{eqn:hrs} and Eq. \ref{eqn:hrs}:
\begin{equation}
\tilde{h(r)} = \left\{
	\begin{array}{ll}
		 x_h h(r) / s_d 	& \quad \text{if $r_n$ is in the varying disc}\\
		 (1-x_h) s(r)/ s_f 	& \quad \text{if $r_n$ is in the hot flow} 
	\end{array} \right. 
\end{equation}
where $\tilde{h(r)}$ indicated the normalized counts in the hard band. Analogous equations for the soft band can be obtained by exchanging $h$ with $s$.\\

\section{Including the QPO}
\label{qpo}

\textsc{propfluc} assumes that the entire hot flow precesses because of frame dragging close to the BH. The precession of the hot flow modulates the emission producing a QPO in the power spectrum and, as described in ID11, ID12, and RIK14, the centroid frequency of the QPO depends both on the radial dimension of the hot flow and on its surface density profile. Modulation occurs mainly through two mechanisms: variation of the projected area of the hot flow towards the line of sight, and variations of the rate at which seed photons coming from the disc enter the hot flow and Compton up-scatter in the optically thin plasma (ID11, ID12, IK13). We can approximate the
former case by multiplying the broad band noise time series with a time series representing the QPO. This is because the observed flux is $\sim L(t) \Omega(t)$, where $L(t)$ is the intrinsic luminosity, which contains the broad band noise, and $\Omega(t)$ is the solid angle subtended by the flow, which is varying quasi-periodically. We can approximate the latter case by adding the broad band noise and QPO time series, since an increase in seed photons adds to the total luminosity available for the hot flow to re-emit. If we make the simplifying assumption that the QPO and broad band noise time series are not correlated with one another and, for the additive case, that
the mean of the QPO time series is zero, we can calculate the resulting power spectrum for both cases analytically (IK13).\\
Now we are fitting also to the cross-spectrum, we must consider how to treat lags intrinsic to the QPO signal. We can define soft and hard band QPO signals, $q_s(t)$ and $q_h(t)$, with Fourier transforms $Q_s(\nu)$ and $Q_h(\nu)$ respectively, such that the cross-spectrum of the QPO signals is $Q_s^*(\nu) Q_h(\nu)=|Q_s(\nu)||Q_h(\nu)| e^{ i \phi(\nu) }$ (where $\phi(\nu)$ is the lag between the soft and hard energy band at frequency $\nu$). For the additive case, the total flux observed in the soft band is $ f_{tot,s}(t) = f_s(t) + q_s(t) $ (with a similar expression for the hard band). Using the above assumptions, the cross-spectrum of the total flux is:
\begin{equation} \label{eqn:cross}
\begin{split}
C(\nu) & = F_{tot,s}^*(\nu) F_{tot,h}(\nu)  \\
           & = F_s^*(\nu) F_h(\nu) + |Q_s(\nu)||Q_h(\nu)| e^{ i \phi(\nu) }
\end{split}
\end{equation}
Following previous versions of \textsc{propfluc} (see IK13), we model $|Q_s(\nu)|$ and $|Q_h(\nu)|$ as a sum of Lorentzian functions, each representing a different harmonic:
\begin{equation}
|Q_{s/h}(\nu)|^2=\sum_{k=1}^4 |Q_{s/h}(k,\nu)|^2
\label{eqn:harmonics}
\end{equation}
where we consider a total of 4 harmonics ($k$ = 1-4), one being a sub-harmonic ($k$ = 4). We then need to make an assumption for the form of the phase, $\phi(\nu)$. The simplest treatment is to also break $\phi(\nu)$ down into harmonics so that we can specify only one model parameter per harmonic, $\phi_k$, in order to characterize the QPO phase lags, rather than an entire (unknown) function, $\phi(\nu)$. Eq. \ref{eqn:harmonics} follows from this because, at the peak frequency of one harmonic in the soft band, the power in all the other harmonics in the hard band is very low (and vice versa). In the additive case, the cross-spectrum becomes:
\begin{equation}
C(\nu) =  F_s^*(\nu) F_h(\nu) + \sum_{k=1}^4 |Q_s(\nu)||Q_h(\nu)| e^{i\phi_k}
\label{eqn:crossadd}
\end{equation}
and we can fit features in the cross-spectrum attributable to the QPO using only one parameter for QPO phase per non-zero harmonic. Note that Eq. \ref{eqn:crossadd} is the equivalent of fitting the real and imaginary part of the cross-spectrum with our broad band noise model plus a sum of Lorentzian functions representing the QPO, with the centroid and widths tied between real and imaginary parts, but with the Lorentzian normalizations free to be different between real and imaginary parts. This is a favorable treatment, since it is the simplest possible way of accounting for phase lags contributed by the QPO, and it is the treatment we adopt in this paper. \\
If we were to instead consider the multiplicative case, equation \ref{eqn:crossadd} picks up an extra term, becoming:
\begin{eqnarray}
\begin{aligned}
C(\nu) = \quad &F_s^*(\nu) F_h(\nu) + \sum_{k=1}^4 |\tilde{Q}_s(k,\nu)||\tilde{Q}_h(k,\nu)| e^{i\phi_k}+ \\
	    &   \tilde{F}_s^*(\nu) \tilde{F}_h(\nu) \otimes \sum_{k=1}^4 |\tilde{Q}_s(k,\nu)||\tilde{Q}_h(k,\nu)| e^{ i \phi_k }
\end{aligned}
\label{eqn:crossmul}
\end{eqnarray}
Here, the $\otimes$ denotes a convolution, and we employ the convention from IK13 that a tilde signifies zero mean in time space and a zero $\nu=0$ component in Fourier space (i.e. $\tilde{x}(t) = x(t) - \langle x(t) \rangle$ or $\tilde{X}(\nu)= X(\nu) - X(0) \delta(\nu)$). Because of the convolution, the third term in this equation can be fairly broad in Fourier space. Therefore the parameters $\phi_k$ can have an influence on frequencies in the cross-spectrum not dominated by the QPO signal. This is best avoided, since our assumptions regarding the QPO are rather \textit{ad hoc}. We note, however, that the first and second terms in Eq. \ref{eqn:crossmul} have amplitudes of $\sim \sigma_{bbn}^2$ (the fractional variability amplitude of the broad band noise) and $\sim \sigma_{qpo}^2$ (the fractional variability of the QPO) respectively, whereas the amplitude of the problematic third term is $\sigma_{bbn}^2 \sigma_{qpo}^2$. Since $\sigma_{bbn}$ and $\sigma_{qpo}$ are both $\sim 0.1$, the third term is small compared to the other terms and so the additive and multiplicative cases are
reasonably similar to one another.

\end{appendix}

{\renewcommand{\arraystretch}{2.0} 

\begin{table*}
\caption{Best fit \textsc{propfluc} parameters (double hump). Errors correspond to 1$\sigma$ confidence level. The subscripts $s$ and $h$ correspond to soft and hard band respectively. The symbol $\sim$ means that the parameter is fixed at the value in column 2 for all the GTIs, the symbol - means that the component did not significantly improve the $\chi^2$ and was omitted, and the symbol $\uparrow$ indicates 3$\sigma$ upper limit (see GTI4). The last row shows the F probability relative to a single hump fit. $T_{d,max}$ and $x_s$ are fixed model parameters computed from previous energy spectral analysis.}
\label{tab:res}

\begin{tabular}{ccccccc}

\hline

MJD & 55464.4093317 [1] & 55464.5541349 [2] & 55465.0086948 [3] & 55465.2095189 [4]& 55465.3433757 [5]& 55465.6217809 [6]\\
\hline\hline

$\Sigma_0$ & 2.75$^{+0.32}_{-0.19}$ & 1.34$^{+0.22}_{-0.25}$ & 6.18$^{+0.64}_{-0.35}$ & 3.35$^{+0.19}_{-0.12}$ & 5.11$^{+0.67}_{-0.55}$ & 2.31$^{+0.49}_{-0.75}$\\
$F_{var} [\%]$ & 34.60$^{+0.11}_{-1.13}$ & 29.46$^{+0.28}_{-0.41}$ & 30.78$^{+0.08}_{-0.11}$ & 31.17$^{+0.56}_{-0.16}$ & 21.30$^{+0.15}_{-4.23}$ & 24.71$^{+0.49}_{-1.43}$\\
$\zeta$ & 0 & $\sim$ & $\sim$ & $\sim$ & $\sim$ & $\sim$\\
$\lambda$ & 0.9 & $\sim$ & $\sim$ & $\sim$ & $\sim$ & $\sim$\\
$\kappa$ & 3.0 & $\sim$ & $\sim$ & $\sim$ & $\sim$ & $\sim$\\
$r_i$ & 4.5 & $\sim$ & $\sim$ & $\sim$ & $\sim$ & $\sim$\\
$r_{bw}$ & 5.0 & $\sim$ & $\sim$ & $\sim$ & $\sim$ & $\sim$\\
$r_o$ & 58.55$^{+0.26}_{-0.17}$ & 54.32$^{+0.15}_{-0.15}$ & 43.18$^{+0.17}_{-0.09}$ & 39.45$^{+0.09}_{-0.09}$ & 37.60$^{+0.06}_{-0.05}$ & 31.05$^{+0.07}_{-0.05}$\\
$N_{var}$ & 0.51$^{+0.12}_{-0.07}$ & 0.50$^{+0.05}_{-0.02}$ & 0.15$^{+0.46}_{-0.46}$ & 0.0002 $\uparrow$ & 1.27$^{+0.34}_{-0.31}$ & 0.84$^{+0.22}_{-0.07}$\\
$\Delta d$ & 35.0 & $\sim$ & $\sim$ & $\sim$ & $\sim$ & $\sim$\\
$\nu_{d,max}$ & 0.074$^{+0.061}_{-0.029}$ & 0.050$^{+0.041}_{-0.021}$ & 0.011$^{+0.124}_{-0.124}$ & 0.027 & 1.182$^{+0.477}_{-0.230}$ & 0.844$^{+0.245}_{-0.060}$\\
$T_{d,max}$ [keV] & 0.13 & 0.13 & 0.18 & 0.20 & 0.22 & 0.28\\
$Q$ & 1.92$^{+0.35}_{-0.11}$ & 2.67$^{+0.14}_{-0.10}$ & 3.79$^{+0.47}_{-0.27}$ & 2.91$^{+0.37}_{-0.15}$ & 4.13$^{+0.24}_{-0.11}$ & 2.97$^{+0.25}_{-0.11}$\\
$Q_{sub}$ & 1.99$^{+1.63}_{-1.15}$ & 2.40$^{+0.76}_{-0.45}$ & 1.09$^{+0.78}_{-0.52}$ & 0.50$^{+0.09}_{-0.04}$ & 0.21$^{+0.08}_{-0.06}$ & 0.58$^{+0.35}_{-0.12}$\\
$\sigma_{QPO,h} [\%]$ & 16.18$^{+0.09}_{-0.53}$ & 16.57$^{+0.10}_{-0.63}$ & 13.89$^{+0.14}_{-0.21}$ & 15.74$^{+0.11}_{-0.22}$ & 13.21$^{+0.08}_{-0.19}$ & 12.58$^{+0.08}_{-0.17}$\\
$\sigma_{QPO2,h} [\%]$ & 5.19$^{+1.02}_{-1.46}$ & 8.03$^{+0.21}_{-0.66}$ & 6.83$^{+1.41}_{-1.09}$ & 6.21$^{+1.10}_{-1.30}$ & 7.14$^{+1.09}_{-0.82}$ & 6.27$^{+0.98}_{-1.34}$\\
$\sigma_{QPO3,h} [\%]$ & - & 7.51$^{+0.37}_{-1.18}$ & 5.68$^{+1.04}_{-1.36}$ & 9.55$^{+0.17}_{-0.77}$ & 6.51$^{+1.05}_{-1.16}$ & 7.52$^{+1.30}_{-0.69}$\\
$\sigma_{QPO_{sub},h} [\%]$ & 2.98$^{+2.55}_{-1.67}$ & 5.92$^{+0.44}_{-0.67}$ & 2.90$^{+2.97}_{-2.15}$ & 7.72$^{+1.81}_{-0.85}$ & 4.04$^{+2.07}_{-1.90}$ & 3.63$^{+1.00}_{-1.63}$\\
$\sigma_{QPO,s} [\%]$ & 13.88$^{+0.12}_{-0.99}$ & 12.01$^{+0.21}_{-0.23}$ & 6.98$^{+0.41}_{-0.20}$ & 4.44$^{+0.22}_{-0.39}$ & 5.10$^{+0.18}_{-0.17}$ & 5.22$^{+0.12}_{-0.14}$\\
$\sigma_{QPO2,s} [\%]$ & 10.17$^{+0.29}_{-0.52}$ & 7.35$^{+0.87}_{-1.15}$ & 4.58$^{+0.38}_{-0.85}$ & 3.27$^{+0.28}_{-1.19}$ & 2.03$^{+0.30}_{-1.17}$ & 1.06$^{+0.14}_{-0.82}$\\
$\sigma_{QPO3,s} [\%]$ & - & 7.55$^{+1.26}_{-1.41}$ & 3.61$^{+0.38}_{-1.12}$ & 2.18$^{+0.33}_{-1.14}$ & 1.91$^{+0.27}_{-1.10}$ & 3.25$^{+0.17}_{-0.48}$\\
$\sigma_{QPO_{sub},s} [\%]$ & 7.80$^{+0.17}_{-0.94}$ & 6.70$^{+0.80}_{-1.13}$ & 5.67$^{+0.70}_{-1.02}$ & 9.69$^{+0.05}_{-0.26}$ & 9.10$^{+0.04}_{-0.47}$ & 5.61$^{+0.38}_{-0.59}$\\
$\phi_{QPO}[cycles]$ & 0.01$^{+0.00}_{-0.00}$ & 0.01$^{+0.00}_{-0.00}$ & 0.02$^{+0.01}_{-0.01}$ & 0.01$^{+0.02}_{-0.01}$ & 0.12$^{+0.01}_{-0.01}$ & 0.11$^{+0.00}_{-0.00}$\\
$\phi_{QPO2}[cycles]$ & -0.01$^{+0.05}_{-0.05}$ & 0.02$^{+0.03}_{-0.02}$ & -0.00$^{+0.05}_{-0.05}$ & -0.04$^{+0.06}_{-0.08}$ & -0.08$^{+0.09}_{-0.14}$ & 0.05$^{+0.32}_{-0.35}$\\
$\phi_{QPO3}[cycles]$ & - & -0.06$^{+0.03}_{-0.03}$ & 0.03$^{+0.07}_{-0.07}$ & 0.10$^{+0.15}_{-0.08}$ & 0.12$^{+0.18}_{-0.23}$ & 0.07$^{+0.05}_{-0.06}$\\

$\phi_{QPO_{sub}}[cycles]$ & 0.05$^{+0.15}_{-0.07}$ & 0.05$^{+0.04}_{-0.02}$ & -0.11$^{+0.11}_{-0.25}$ & 0.06$^{+0.03}_{-0.01}$ & -0.10$^{+0.20}_{-0.17}$ & -0.12$^{+0.17}_{-0.13}$\\

$\gamma_s$ & 3.0 & $\sim$ & $\sim$ & $\sim$ & $\sim$ & $\sim$\\
$\gamma_h$ & 4.5 & $\sim$ & $\sim$ & $\sim$ & $\sim$ & $\sim$\\
$x_s$ & 0.16 & 0.17 & 0.26 & 0.35 & 0.38 & 0.50\\
$\chi^2/dof$ & 550.50/380 & 639.57/377 & 502.64/377 & 646.89/377 & 508.49/377 & 497.85/377 \\
$P_f[\%]$ & 100.00 & 5.04 & 67.99 & 100.0 & 2.85 & 0.03 \\
\end{tabular}
\end{table*}

\begin{table*}
\renewcommand\thetable{1}
\caption{Continued.}
\label{}

\begin{tabular}{cccccc}

\hline

MJD & 55465.8117549 [7] & 55466.1446146 [8] & 55466.2118039 [9]& 55466.34714 [10]& 55467.4190101 [11]\\
\hline\hline

$\Sigma_0$ & 1.19$^{+0.24}_{-0.27}$ & 7.33$^{+0.45}_{-0.28}$ & 6.35$^{+0.88}_{-0.52}$ & 4.31$^{+0.91}_{-0.53}$ & 4.19$^{+1.09}_{-0.77}$\\
$F_{var} [\%]$ & 25.97$^{+0.76}_{-1.40}$ & 17.41$^{+1.34}_{-2.14}$ & 31.96$^{+0.43}_{-0.16}$ & 30.03$^{+11.45}_{-1.44}$ & 28.20$^{+0.35}_{-0.70}$\\
$\zeta$ & 0 & $\sim$ & $\sim$ & $\sim$ & $\sim$\\
$\lambda$ & 0.9 & $\sim$ & $\sim$ & $\sim$ & $\sim$\\
$\kappa$ & 3.0 & $\sim$ & $\sim$ & $\sim$ & $\sim$\\
$r_i$ & 4.5 & $\sim$ & $\sim$ & $\sim$ & $\sim$\\
$r_{bw}$ & 5.0 & $\sim$ & $\sim$ & $\sim$ & $\sim$\\
$r_o$ & 29.23$^{+0.07}_{-0.08}$ & 28.47$^{+0.04}_{-0.04}$ & 27.78$^{+0.12}_{-0.07}$ & 27.32$^{+0.12}_{-0.12}$ & 20.65$^{+0.04}_{-0.04}$\\
$N_{var}$ & 1.09$^{+0.01}_{-0.06}$ & 1.38$^{+0.37}_{-0.36}$ & 0.23$^{+0.01}_{-0.01}$ & 0.44$^{+0.09}_{-0.03}$ & 0.50$^{+0.04}_{-0.02}$\\
$\Delta d$ & 35.0 & $\sim$ & $\sim$ & $\sim$ & $\sim$\\
$\nu_{v,max}$ & 1.113$^{+0.052}_{-0.070}$ & 2.017$^{+0.248}_{-0.217}$ & 0.054$^{+0.035}_{-0.012}$ & 0.389$^{+0.295}_{-0.111}$ & 1.232$^{+0.248}_{-0.109}$\\
$T_{d,max}$ [keV] & 0.35 & 0.26 & 0.27 & 0.29 & 0.38\\
$Q$ & 5.08$^{+0.87}_{-0.42}$ & 10.30$^{+2.70}_{-1.49}$ & 6.53$^{+2.44}_{-1.31}$ & 6.81$^{+1.39}_{-0.67}$ & 9.83$^{+4.49}_{-2.57}$\\
$Q_{sub}$ & 6.77$^{+3.39}_{-3.39}$ & 0.16$^{+0.07}_{-0.04}$ & 0.63$^{+0.08}_{-0.05}$ & 0.59$^{+0.38}_{-0.15}$ & 30.73$^{+0.01}_{-0.01}$\\
$\sigma_{QPO,h} [\%]$ & 10.51$^{+0.16}_{-0.18}$ & 8.61$^{+0.88}_{-0.37}$ & 8.89$^{+0.94}_{-0.25}$ & 9.68$^{+1.02}_{-0.25}$ & 7.25$^{+0.39}_{-0.14}$\\
$\sigma_{QPO2,h} [\%]$ & 6.77$^{+0.60}_{-0.49}$ & 2.80$^{+0.54}_{-2.43}$ & 4.04$^{+0.73}_{-2.07}$ & 4.97$^{+1.24}_{-1.60}$ & 6.52$^{+0.44}_{-0.31}$\\
$\sigma_{QPO3,h} [\%]$ & 4.39$^{+0.57}_{-1.18}$ & - & - & 3.02$^{+1.82}_{-1.82}$ & -\\
$\sigma_{QPO_{sub},h} [\%]$ & 3.45$^{+0.58}_{-1.23}$ & 13.19$^{+0.18}_{-2.90}$ & 6.09$^{+2.00}_{-1.13}$ & 2.99$^{+0.25}_{-1.55}$ & 2.49$^{+0.20}_{-0.63}$\\
$\sigma_{QPO,s} [\%]$ & 4.17$^{+0.09}_{-0.17}$ & 3.23$^{+0.07}_{-0.18}$ & 2.85$^{+0.10}_{-0.35}$ & 3.32$^{+0.56}_{-0.31}$ & 2.23$^{+0.06}_{-0.21}$\\
$\sigma_{QPO2,s} [\%]$ & 2.51$^{+0.12}_{-0.72}$ & - & 0.75$^{+0.15}_{-0.15}$ & 1.66$^{+0.22}_{-0.22}$ & 0.75$^{+0.08}_{-0.53}$\\
$\sigma_{QPO3,s} [\%]$ & 3.61$^{+0.17}_{-0.73}$ & - & - & - & - \\
$\sigma_{QPO_{sub},s} [\%]$ & 2.17$^{+0.36}_{-0.64}$ & 6.36$^{+0.54}_{-1.12}$ & 7.93$^{+0.57}_{-0.22}$ & 7.71$^{+1.31}_{-0.38}$ & 0.33$^{+0.04}_{-0.04}$\\
$\phi_{QPO}[cycles]$ & 0.08$^{+0.01}_{-0.01}$ & 0.16$^{+0.01}_{-0.01}$ & 0.14$^{+0.01}_{-0.02}$ & 0.15$^{+0.01}_{-0.01}$ & 0.12$^{+0.01}_{-0.02}$\\
$\phi_{QPO2}[cycles]$ & -0.02$^{+0.05}_{-0.05}$ & - & -0.09$^{+0.01}_{-0.01}$ & -0.01$^{+0.11}_{-0.11}$ & 0.05$^{+0.26}_{-0.13}$\\
$\phi_{QPO3}[cycles]$ & 0.07$^{+0.07}_{-0.05}$  & - & - & - & - \\

$\phi_{QPO_{sub}}[cycles]$ & -0.07$^{+0.07}_{-0.07}$ & 0.05$^{+0.09}_{-0.01}$ & -0.01$^{+0.04}_{-0.04}$ & 0.04$^{+0.15}_{-0.17}$ & 0.07$^{+0.01}_{-0.01}$\\

$\gamma_s$ & 3.0 & $\sim$ & $\sim$ & $\sim$ & $\sim$\\
$\gamma_h$ & 4.5 & $\sim$ & $\sim$ & $\sim$ & $\sim$\\
$x_s$ & 0.61 & 0.48 & 0.48 & 0.51 & 0.62\\
$\chi^2/dof$ & 512.15/377 & 522.29/383 & 519.55/380 & 488.53/379 & 558.51/381 \\
$P_f[\%]$ & $<$ 0.01 & 0.51 & 0.20 & 0.02 & $<$ 0.01\\
\end{tabular}
\end{table*}

\begin{table*}
\caption{Summary of new model parameters.}
\label{tab:pars}
\begin{tabular}{ccccc}
 \hline

 & Parameter & Description 

 \\
 \hline
 \hline
1  & $r_d$    & Disc radius in units of $R_g$ \\
 \hline
2  & $\Delta d$       & radial extension of the disc in units of $R_g$ \\
 \hline
3  & $N_{var}$      & fraction of hot flow variability in the disc at the truncation radius  \\
 \hline
4  & $T_{d,max}$     & maximum temperature in the disc [keV] \\
 \hline
5  & $\nu_{d,max}$       & maximum viscous frequency in the disc [Hz]\\
 \hline
6  & $x_s$      & fraction of disc emission in the soft band \\
 \hline
7  & $n_h$     & hydrogen column density [$10^{22} cm^{-2}$] \\
 \hline
8  & $\phi_{qpo}$        & main QPO phase lag [cycles]   \\
 \hline
9  &  $\phi_{qpo2}$   & QPO second harmonic phase lag [cycles] \\
 \hline
10 &  $\phi_{qpo3}$   & QPO third harmonic phase lag [cycles] \\
 \hline
11 &  $\phi_{qpos}$   & QPO sub-harmonic phase lag [cycles] \\

\hline
\end{tabular}

\label{tab:paras}
\end{table*}  

\label{lastpage}

\begin{thebibliography}{99}

\bibitem[Altamirano \& Strohmayer(2012)]{2012ApJ...754L..23A} Altamirano, D., \& Strohmayer, T.\ 2012, ApJ, 754, L23 

\bibitem[\protect\citeauthoryear{Ar{\'e}valo 
\& Uttley}{2006}]{2006MNRAS.367..801A} Ar{\'e}valo P., Uttley P., 2006, MNRAS, 367, 801 

\bibitem[Asplund et 
al.(2009)]{2009ARA&A..47..481A} Asplund, M., Grevesse, N., Sauval, A.~J., \& Scott, P.\ 2009, ARA\&A, 47, 481 

\bibitem[Balucinska-Church 
\& McCammon(1992)]{1992ApJ...400..699B} Balucinska-Church, M., \& McCammon, D.\ 1992, ApJ, 400, 699 

\bibitem[Belloni et 
al.(1997)]{1997A&A...322..857B} Belloni, T., van der Klis, M., Lewin, W.~H.~G., et al.\ 1997, A\&A, 322, 857 

\bibitem[Belloni et al.(2002)]{2002ApJ...572..392B} Belloni, T., Psaltis, D., \& van der Klis, M.\ 2002, ApJ, 572, 392

\bibitem[Belloni et 
al.(2005)]{2005A&A...440..207B} Belloni, T., Homan, J., Casella, P., et al.\ 2005, A\&A, 440, 207 

\bibitem[\protect\citeauthoryear{Belloni}{2010}]{2010LNP...794...53B} 
Belloni T.~M., 2010, LNP, 794, 53 

\bibitem[B{\"o}ttcher \& Liang(1999)]{1999ApJ...511L..37B} B{\"o}ttcher, M., \& Liang, E.~P.\ 1999, ApJ, 511, L37 

\bibitem[\protect\citeauthoryear{Burrows et al.}{2005}]{2005SSRv..120..165B} Burrows, D.~N., Hill, 
J.~E., Nousek, J.~A., et al.\ 2005, Space Sci. Rev., 120, 165 

\bibitem[\protect\citeauthoryear{Casella}{2005}]{2005ApJ...629..403C}
Casella P., Belloni T., Stella L., 2005, ApJ, 629, 403

\bibitem[Churazov et al.(2001)]{2001MNRAS.321..759C} Churazov, E., 
Gilfanov, M., \& Revnivtsev, M.\ 2001, MNRAS, 321, 759 

\bibitem[\protect\citeauthoryear{Done, Gierli{\'n}ski, 
\& Kubota}{2007}]{2007A&ARv..15....1D} Done C., Gierli{\'n}ski M., Kubota A.,
2007, A\&ARv, 15, 1 

\bibitem[Ebisawa et al.(1994)]{1994PASJ...46..375E} Ebisawa, K., Ogawa, M., 
Aoki, T., et al.\ 1994, PASJ, 46, 375

\bibitem[Esin et al.(1997)]{1997ApJ...489..865E} Esin, A.~A., McClintock, 
J.~E., \& Narayan, R.\ 1997, ApJ, 489, 865 

\bibitem[Evans et 
al.(2007)]{2007A&A...469..379E} Evans, P.~A., Beardmore, A.~P., Page, K.~L., et al.\ 2007, A\&A, 469, 379 

\bibitem[Fragile et al.(2007)]{2007ApJ...668..417F} Fragile, P.~C., Blaes, 
O.~M., Anninos, P., \& Salmonson, J.~D.\ 2007, ApJ, 668, 417 

\bibitem[\protect\citeauthoryear{Frank, King, 
\& Raine}{2002}]{2002apa..book.....F} Frank J., King A., Raine D.~J.,
Accretion power in astrophysics, 3rd edition, 2002, Cambridge University Press

\bibitem[Gehrels et al.(2004)]{2004ApJ...611.1005G} Gehrels, N., 
Chincarini, G., Giommi, P., et al.\ 2004, ApJ, 611, 1005 

\bibitem[\protect\citeauthoryear{Gilfanov}{2010}]{2010LNP...794...17G} 
Gilfanov M., 2010, LNP, 794, 17

\bibitem[Homan et al.(2001)]{2001ApJS..132..377H} Homan, J., Wijnands, R., 
van der Klis, M., et al.\ 2001, ApJS, 132, 377 

\bibitem[\protect\citeauthoryear{Ingram 
\& Done}{2011}]{2011MNRAS.415.2323I} Ingram A., Done C., 2011 (ID11), MNRAS,
415, 2323

\bibitem[\protect\citeauthoryear{Ingram 
\& Done}{2012}]{2012MNRAS.419.2369I} Ingram A., Done C., 2012 (ID12),
MNRAS, 419, 2369

\bibitem[\protect\citeauthoryear{Ingram 
\& van der Klis}{2013}]{2013MNRAS.434.1476I} Ingram A., van der Klis M., 2013 (IK13),
MNRAS, 434, 1476

\bibitem[Jahoda et al.(1996)]{1996SPIE.2808...59J} Jahoda, K., Swank, 
J.~H., Giles, A.~B., et al.\ 1996, Proc. SPIE, 2808, 59 

\bibitem[Kalamkar et al.(2011)]{2011ApJ...731L...2K} Kalamkar, M., Homan, 
J., Altamirano, D., et al.\ 2011, ApJ, 731, L2 

\bibitem[Kalamkar et al.(2013)]{2013ApJ...766...89K} Kalamkar, M., van der 
Klis, M., Uttley, P., Altamirano, D., \& Wijnands, R.\ 2013, ApJ, 766, 89 

\bibitem[Kalamkar et al.(cd ..)]{2015ApJ...808..144K} Kalamkar, M., van der 
Klis, M., Heil, L., \& Homan, J.\ 2015a, ApJ, 808, 144 


\bibitem[Kalamkar et al.(2015)]{2015ApJ...802...23K} Kalamkar, M., 
Reynolds, M.~T., van der Klis, M., Altamirano, D., 
\& Miller, J.~M.\ 2015b, ApJ, 802, 23 

\bibitem[Kalberla et 
al.(2005)]{2005A&A...440..775K} Kalberla, P.~M.~W., Burton, W.~B., Hartmann, D., et al.\ 2005, A\&A, 440, 775 

\bibitem[Kotov et al.(2001)]{2001MNRAS.327..799K} Kotov, O., Churazov, E., \& Gilfanov, M.\ 2001, MNRAS, 327, 799 

\bibitem[Kuulkers et 
al.(2013)]{2013A&A...552A..32K} Kuulkers, E., Kouveliotou, C., Belloni, T., et al.\ 2013, A\&A, 552, A32 

\bibitem[Leahy et al.(1983)]{1983ApJ...266..160L} Leahy, D.~A., Darbro, W., 
Elsner, R.~F., et al.\ 1983, ApJ, 266, 160 

\bibitem[Lyubarskii(1997)]{1997MNRAS.292..679L} Lyubarskii, Y.~E.\ 1997, 
MNRAS, 292, 679 

\bibitem[Mangano et al.(2010)]{2010GCN..11296...1M} Mangano, V., Hoversten, 
E.~A., Markwardt, C.~B., et al.\ 2010, GRB Coordinates Network, 11296, 1 

\bibitem[Melia 
\& Misra(1993)]{1993ApJ...411..797M} Melia, F., \& Misra, R.\ 1993, ApJ, 411, 797 

\bibitem[Merloni et al.(2000)]{2000MNRAS.313..193M} Merloni, A., Fabian, 
A.~C., \& Ross, R.~R.\ 2000, MNRAS, 313, 193 

\bibitem[Misra(2000)]{2000ApJ...529L..95M} Misra, R.\ 2000, ApJ, 529, L95 

\bibitem[Mitsuda et al.(1984)]{1984PASJ...36..741M} Mitsuda, K., Inoue, H., 
Koyama, K., et al.\ 1984, PASJ, 36, 741 

\bibitem[Miyamoto et al.(1988)]{1988Natur.336..450M} Miyamoto, S., Kitamoto, S., Mitsuda, K., \& Dotani, T.\ 1988, Nature, 336, 450 

\bibitem[Miyamoto \& Kitamoto(1989)]{1989Natur.342..773M} Miyamoto, S., \& Kitamoto, S.\ 1989, Nature, 342, 773

\bibitem[Mu{\~n}oz-Darias et al.(2011)]{2011MNRAS.415..292M} 
Mu{\~n}oz-Darias, T., Motta, S., Stiele, H., 
\& Belloni, T.~M.\ 2011, MNRAS, 415, 292 

\bibitem[Negoro et al.(2010)]{2010ATel.2873....1N} Negoro, H., Yamaoka, K., 
Nakahira, S., et al.\ 2010, The Astronomer's Telegram, 2873, 1 

\bibitem[Nowak et al.(1999)]{1999ApJ...517..355N} Nowak, M.~A., Wilms, J., \& Dove, J.~B.\ 1999a, ApJ, 517, 355 

\bibitem[Nowak et al.(1999)]{1999ApJ...515..726N} Nowak, M.~A., Wilms, J., Vaughan, B.~A., Dove, J.~B., \& Begelman, M.~C.\ 1999b, ApJ, 515, 726 

\bibitem[\protect\citeauthoryear{Psaltis, Belloni, 
\& van der Klis}{1999}]{1999ApJ...520..262P} Psaltis D., Belloni T.,
van der Klis M., 1999, ApJ, 520, 262 

\bibitem[Rapisarda et al.(2014)]{2014MNRAS.440.2882R} Rapisarda, S., 
Ingram, A., \& van der Klis, M.\ 2014 (RIK14), MNRAS, 440, 2882 

\bibitem[Remillard 
\& McClintock(2006)]{2006ARA&A..44...49R} Remillard, R.~A., \& McClintock, J.~E.\ 2006, ARA\&A, 44, 49 

\bibitem[Reynolds 
\& Miller(2013)]{2013ApJ...769...16R} Reynolds, M.~T., \& Miller, J.~M.\ 2013, ApJ, 769, 16 

\bibitem[\protect\citeauthoryear{Shakura 
\& Sunyaev}{1973}]{1973A&A....24..337S} Shakura N.~I., Sunyaev R.~A., 1973,
A\&A, 24, 337 

\bibitem[Stella 
\& Vietri(1998)]{1998ApJ...492L..59S} Stella, L., \& Vietri, M.\ 1998, ApJ, 492, L59 

\bibitem[Sunyaev 
\& Truemper(1979)]{1979Natur.279..506S} Sunyaev, R.~A., \& Truemper, J.\ 1979, Nature, 279, 506 

\bibitem[Svensson 
\& Zdziarski(1994)]{1994ApJ...436..599S} Svensson, R., \& Zdziarski, A.~A.\ 1994, ApJ, 436, 599 

\bibitem[Thorne 
\& Price(1975)]{1975ApJ...195L.101T} Thorne, K.~S., \& Price, R.~H.\ 1975, ApJ, 195, L101

\bibitem[Titarchuk(1994)]{1994ApJ...434..570T} Titarchuk, L.\ 1994, ApJ, 
434, 570 

\bibitem[Uttley 
\& McHardy(2001)]{2001MNRAS.323L..26U} Uttley, P., \& McHardy, I.~M.\ 2001, MNRAS, 323, L26 

\bibitem[Uttley et al.(2005)]{2005MNRAS.359..345U} Uttley, P., McHardy, 
I.~M., \& Vaughan, S.\ 2005, MNRAS, 359, 345 

\bibitem[van der Klis et al. (1995)]{} van der Klis, M. 1995, in \textsc{The lives of Neutron stars}, ed. M. A. Alpar,
U. Kiziloglu, \& J. van Paradijs, 301

\bibitem[Wijnands 
\& van der Klis(1998)]{1998ApJ...507L..63W} Wijnands, R., \& van der Klis, M.\ 1998, ApJ, 507, L63 

\bibitem[Wilkinson 
\& Uttley(2009)]{2009MNRAS.397..666W} Wilkinson, T., \& Uttley, P.\ 2009, MNRAS, 397, 666 

\end{thebibliography}
\end{document}